\documentclass[useAMS,usenatbib]{mn2e}
\usepackage{amssymb}
\usepackage{amsmath}
\usepackage{graphicx}
\usepackage{lscape}
\usepackage{color}
\usepackage{subfigure}

\def\h2{\rm{H_2}}
\def\fh2{f_{\rm{H_2}}}
\def\SHI{\Sigma_{\rm{HI}}}
\def\Sh2{\Sigma_{\h2}}
\def\sgas{\Sigma_{\rm{gas}}}
\def\scrit{\Sigma_{\rm{crit}}}
\def\zgas{\rm{[Z/H]_{gas}}}
\def\ms{M_{\odot}}
\def\Hs{M_{\h2}/M_*}
\def\HIs{M_{\rm{HI}}/M_*}
\def\HHI{M_{\h2}/M_{\rm{HI}}}
\def\mspc{M_{\odot}\rm{pc^{-2}}}
\def\mugas{\mu_{\rm{gas}}}

\title[HI to $H_2$ Transition and Scaling Properties]
{The Atomic to Molecular Transition and its Relation to the Scaling Properties of Galaxy Disks in the Local Universe}

\author[Fu et al.]{Jian Fu$^{1,2,3}$ \thanks{E-mail: fujian@shao.ac.cn; fujian@mpa-garching.mpg.de},
Qi Guo$^{4,2}$, Guinevere Kauffmann$^{2}$, Mark R. Krumholz$^{5}$\\
$^1$Key Laboratory for Research in Galaxies and Cosmology, Shanghai Astronomical Observatory, CAS,\\ 80 Nandan Rd., Shanghai, 200030, China \\
$^2$Max-Planck-Institut f\"ur Astrophysik, D-85740 Garching, Germany  \\
$^3$Graduate School, the Chinese Academy of Sciences, Beijing, 100039, China \\
$^4$Institute for Computational Cosmology, Physics Department, Durham, U.K. \\
$^5$Department of Astronomy and Astrophysics, University of California, Santa Cruz, CA 95064, USA }

\begin{document}

\maketitle

\begin{abstract}

We extend existing semi-analytic models of galaxy formation to track atomic and molecular gas in disk galaxies. Simple recipes for
processes such as cooling, star formation, supernova feedback, and chemical enrichment of the stars and gas are grafted on to dark
matter halo merger trees derived from the Millennium Simulation. Each galactic disk is represented by a series of concentric rings. We
assume that surface density profile of infalling gas in a dark matter halo is exponential, with scale radius $r_{\rm d}$ that is
proportional to the virial radius of the halo times its spin parameter $\lambda$. As the dark matter haloes grow through mergers and
accretion, disk galaxies assemble from the inside out. We include two simple prescriptions for molecular gas formation processes in
our models: one is based on the analytic calculations by Krumholz, McKee \& Tumlinson (2008), and the other is a prescription where
the $\rm{H_2}$ fraction is determined by the pressure of the interstellar medium (ISM). Motivated by the observational results of
Leroy et al. (2008), we adopt a star formation law in which $\Sigma_{\rm{SFR}}\propto\Sigma_{\rm{H_2}}$ in the regime where the
molecular gas dominates the total gas surface density, and $\Sigma_{\rm{SFR}}\propto \Sigma_{\rm gas}^2$ where atomic hydrogen
dominates. We then fit these models to the radial surface density profiles of stars, HI and $\rm{H_2}$ drawn from recent high
resolution surveys of stars and gas in nearby galaxies. We explore how the ratios of atomic gas, molecular gas and stellar mass vary
as a function of global galaxy scale parameters, including stellar mass, stellar surface density, and gas surface density. We
elucidate how the trends can be understood in terms of three variables that determine the partition of baryons in disks: the mass of
the dark matter halo, the spin parameter of the halo, and the amount of gas recently accreted from the external environment.

\end{abstract}

\begin{keywords}
galaxies: evolution - stars: formation - galaxies: ISM - ISM: atoms - ISM: molecules
\end{keywords}

\section{Introduction}

Before we can reliably compute how galaxies form stars and evolve as a function of cosmic time, we must understand the physical
processes that regulate the balance between neutral and molecular gas in their interstellar media. Only if $\h2$ forms, will
gravitationally unstable clouds cool and collapse to high enough densities to trigger star formation in the first galaxies. It is also
generally believed that star formation occurs exclusively in molecular clouds in all galaxies at all epochs.

Many galaxy formation models adopt the so-called ``Kennicutt-Schmidt'' law (hereafter K-S law, Schmidt 1959, Kennicutt 1998) to
prescribe the rate at which a disk galaxy of given cold gas mass and scale radius will form its stars. This has the form
\begin{equation}\label{eq:kslaw}
\Sigma_{\rm{SFR}}\propto\sgas^{n}
\end{equation}
where $\Sigma_{\rm{SFR}}$ represents the star formation rate surface density, $\sgas$ is the total surface density of the cold gas in
the disk, and the exponent $n=1.4$ is often adopted. Some semi-analytic models also account for a critical density below which disks
become gravitationally stable and star formation no longer occurs (e.g Kauffmann 1996, De Lucia \& Blaizot 2007). In this case,
\begin{equation}\label{eq:kslaw1}
\Sigma_{\rm{SFR}}\propto[\sgas-\scrit]
\end{equation}
where the critical density $\scrit$ is evaluated using the disk stability criterion given in Toomre (1964). In both Equations
(\ref{eq:kslaw}) and (\ref{eq:kslaw1}), the star formation rate surface density is proportional to the total surface density of cold
gas (i.e. both HI and H$_2$ components) in the galaxy. This prescription was motivated by the analysis of 97 nearby galaxies by
Kennicutt (1998), which showed that star formation is more tightly correlated with $\sgas$ than with $\Sigma_{\h2}$. There have been
studies in apparent disagreement with these conclusions; for example, Wong \& Blitz (2002) found that the relation between
$\Sigma_{\rm{SFR}}$ and $\Sigma_{\h2}$ is stronger than that between $\Sigma_{\rm{SFR}}$ and $\Sigma_{\rm{gas}}$ in galaxies with high
molecular gas fractions. In recent years, high quality, spatially-resolved maps of the cold gas have become available for samples of a
few dozen nearby galaxies. Examples of such data include HI maps from The HI Nearby Galaxy Survey (THINGS) and CO maps from the
Berkeley-Illinois-Maryland Association Survey of Nearby Galaxies (BIMA SONG) and HERA CO-Line Extragalactic Survey (HERACLES).
Measurements of the rate at which stars are forming at different radii in the galaxy are provided by Spitzer and GALEX observations.
The combination of these different data sets has led to important new constraints on the relationship between star formation and gas
in galactic disks. Bigiel et al. (2008) studied 18 disk galaxies and showed that $\rm{H_2}$ forms stars at a roughly constant
efficiency in spirals at radii where it can be detected. Their results suggest a star formation law of the form
\begin{equation}\label{eq:kslawh2}
\Sigma_{\rm{SFR}}\propto\Sigma_{\rm{H_2}}^{1.0\pm0.2}
\end{equation}

Motivated by these findings, galaxy formation modelers are now progressing beyond a simple single-component view of the
cold phase of the interstellar medium, and are attempting to model the formation of molecular hydrogen in galaxies.
Gnedin, Tassis \& Kravtsov (2009) included a phenomenological model for $\h2$ formation in hydrodynamic simulations of
disk galaxy formation. Their model includes nonequilibrium formation of $\h2$ on dust and approximate treatment of both
its self-shielding, and shielding by dust from the dissociating UV radiation field. Dutton (2009) and Dutton \& van den
Bosch (2009) utilized the empirically-motivated hypothesis of Blitz and Rosolowsky (2004,2006) that hydrostatic
pressure alone determines the ratio of atomic to molecular gas averaged over a particular radius in the disk in their
analytic models of disk formation in a $\Lambda$CDM cosmology. They analyzed the radial distribution of stars and star
formation in their disks, but did not focus very much on gas properties in their model. There have also been some
attempts to predict the balance between atomic and molecular gas in galaxies at different redshifts by post-processing
the publically available outputs of semi-analytic galaxy formation models (e.g. Obreschkow et al. 2009). This work also
used the same Blitz and Rosolowsky (2004, 2006) prescription to predict the fraction of molecular gas in disks.
However, the Obreschkow et al. approach is not self-consistent, because the simulations have been run assuming a
``standard'' Kennicutt-Schmidt law for star formation and the presence or absence of molecular gas has no influence on
the actual evolution of the galaxies in the model.

In this paper, we develop new semi-analytic models that follow gas cooling, supernova feedback, the assembly of
galactic disks, the conversion of atomic gas into molecular gas as a function of radius within the disk, and the
conversion of the gas into stars. In the 1990's, semi-analytic models of galaxy formation were developed into a useful
technique for interpreting observational data on galaxy populations (e.g. Kauffmann, White \& Guiderdoni 1993; Cole et
al. 1994; Somerville \& Primack 1999). In the first decade of the new Millennium, considerable effort went into
grafting these models on to large N-body simulations of the dark matter component of the Universe. These efforts began
with relatively low resolution simulations (Kauffmann et al. 1999), but have rapidly progressed to simulations with
high enough resolution to follow the detailed assembly histories of millions of galaxies with luminosities well below
$L_*$ (Croton et al. 2006; Bower et al. 2006; De Lucia \& Bliazot 2007; Guo et al. 2010).

Our new models are an extension of the techniques described in Croton et al. (2006) and De Lucia \& Blaizot (2007) and
are implemented using the merger trees from the Millennium Simulation (Springel et al. 2005). We explore two different
``recipes'' for partitioning the cold gas into atomic and molecular form: a) a prescription based on the analytic
models of $\h2$ formation, dissociation and shielding developed by Krumholz, McKee \& Tumlinson (2009), in which the
molecular fraction is a local function of the surface density and the metallicity of the cold gas, b) the same
pressure-based formulation explored by Obreschkow et al. (2009).

We first use our models to calculate the HI, $\h2$, stellar mass and SFR surface density profiles of disk galaxies that form in dark
matter haloes with circular velocities $v_{\rm{cir}} \sim$ 200 km/s (i.e. galaxies comparable to the Milky Way) and we compare our
results to the THINGS/HERACLES observations presented in Bigiel et al. (2008).

We then turn to the issue of the predicted {\em scaling relations} between atomic gas, molecular gas and stars for an ensemble of disk
galaxies forming in dark matter haloes spanning a range of different circular velocities. We currently enjoy a rich and diverse array
of scaling laws that describe the stellar components of galaxies. For example, the Tully-Fisher relation and the size-mass relation
for local spiral galaxies play a crucial role in constraining current theories of disk galaxy formation. Likewise, the scaling laws of
bulge-dominated galaxies (the Fundamental Plane) provide important constraints on how these systems may have assembled through
merging. In contrast, few well-established scaling laws exist describing how the cold gas is correlated with other global physical
properties of galaxies. Surveys of atomic and molecular gas in well-defined samples of a few hundred to a thousand galaxies are
currently underway, and this paper will explore what can be learned about disk galaxy formation from the results.

Our paper is organized as follows. In section 2, we briefly describe the simulation used in our study as well as the semi-analytic
model used to track the formation of galaxies in the simulation. In section 3, we describe the new aspects of the models presented in
this paper, including our spatially resolved treatment of disk formation in radial bins, the recipes that prescribe how atomic gas is
converted into molecular gas, and our new prescriptions for star formation and feedback. In section 4, we compare the radial profiles
in our models to observations from the THINGS/HERACLES surveys, and present the global gas properties of the galaxies in our model,
such as atomic and molecular gas mass functions. In section 5, we introduce a set of scaling relations for the atomic and molecular
gas fractions of galaxies and we clarify which aspects of the input physics are responsible for setting the slope and the scatter of
these relations. Finally, in section 6 we summarize our work and discuss our findings.

\section{The simulation and semi-analytic model}

In this section, we give a brief description of the Millennium Simulation and the physical processes treated in the semi-analytic
galaxy formation code L-Galaxies. In the next section, we describe our own changes to the code, which include a resolved model for
disk assembly and new recipes to treat molecular gas formation, star formation and supernova feedback.

\subsection{Mass skeleton: the Millennium simulation}

The Millennium simulation (Springel et al. 2005) is a very large N-body cosmological simulation with N=$2160^3\approx10^{10}$
collisionless particles in a comoving box of 500$h^{-1}$Mpc on a side. The mass of each particle is $8.6 \times {10^8}{M_ \odot }{h^{
- 1}}$. The cosmogony is $\rm{\Lambda}$CDM with parameters $\Omega_\Lambda=0.75,~\Omega_m= 0.25,~\Omega_{\rm{baryon}}=0.045,~\sigma_8=
0.9$ and $h=0.73$. The outputs of the Millennium Simulation are stored in a series of 64 snapshots (0-63); the redshifts of snapshots
4 to 63 are given by the expression
\begin{equation}\label{eq:snapshot}
z_n=10^{(n-63)(n-28)/4200}-1~~~~~~~n = 4,5...63
\end{equation}
The redshifts for snapshots 0 to 3 are $z_0=127$, $z_1=80$, $z_2=50$, $z_3=30$ respectively. Snapshot 63 corresponds to redshift 0 and
the time interval between two snapshots is approximately 200 Myr.

In addition to the main Millennium simulation, there is a smaller version with the same cosmological parameters and mass resolution,
the so-called milli-Millennium, which includes N=$270^3$ particles in a comoving box of 62.5$h^{-1}$Mpc on a side. The
milli-Millennium simulation has the same resolution to the full Millennium simulation and it is very useful for fast exploration of
parameter space. The full simulation is needed to build up sufficient statistics to accurately characterize the distributions of
galaxy properties in a multi-dimensional parameter space.

\subsection{The semi-analytic model: L-Galaxies}

L-Galaxies is the semi-analytic code written by Volker Springel, described in detail in Croton et al. (2006) and updated in De Lucia
\& Blaizot (2007) and subsequent papers by the Munich group. It operates on the dark matter halo and subhalo merger trees constructed
from the 64 snapshots of the Millennium Simulation and specifies the treatment of the following physical processes: reionization, gas
infall and cooling, star formation and metal production, supernova feedback, galaxy mergers and star bursts, black hole growth and AGN
feedback. Here we will briefly review those aspects of the models most relevant to the analysis presented in this paper.

As described above, the time interval between two successive snapshots from the Millennium Run is about 200 Myr. In order to model the
cooling and star formation accurately, the time interval between snapshots is divided into 20 time steps of around 10 Myr, a time
interval that is well matched to the evolutionary time-scale of massive stars. All the physical processes treated by L-Galaxies are
computed at each time step, but the properties of dark matter haloes are only updated at the beginning of each snapshot.

The gas cooling processes in L-Galaxies follow the treatment first outlined in White \& Frenk (1991). In each dark matter halo, the
hot gas is assume to be distributed isothermally with a density profile
\begin{equation}\label{eq:rhohot}
{\rho _{\rm{g}}}\left( r \right) = \frac{{{m_{{\rm{hot}}}}}}{{4\pi {R_{{\rm{vir}}}}{r^2}}},
\end{equation}
where $R_{\rm{vir}}$ is the virial radius of a halo, defined as the radius within which the dark matter density is 200 times the
critical density, and $m_{\rm{hot}}$ is the mass of hot gas within $R_{\rm{vir}}$. The local cooling time of hot gas is the ratio of
the specific thermal energy to the cooling rate per unit volume
\begin{equation}\label{eq:coolingtime}
t_{\rm{cool}}\left(r\right)=\frac{3\bar \mu m_{\rm{p}}k_{\rm{B}}T}{2{\rho _{\rm{g}}\left(r\right)\Lambda\left( {T,Z}
\right)}}
\end{equation}
where $\bar \mu m_{\rm{p}}$ is the mean particle density, $k_{\rm{B}}$ is the Boltzmann constant, and $\Lambda\left(T,Z\right)$ is the
cooling rate. $\Lambda\left(T,Z\right)$ is dependent on the metallicity and the virial temperature $T = 35.9{\left(
{{V_{{\rm{vir}}}}/{\rm{km~s}^{ - 1}}} \right)^2}{\rm{K}}$ of the hot halo gas. We adopt the cooling function computed by Sutherland \&
Dopita (1993) and the cooling radius $r_{\rm{cool}}$ is defined as the radius where $t_{\rm{cool}}$ is equal to the dynamical time of
the halo $t_{\rm{dyn}}^{\rm{halo}}= R_{{\rm{vir}}}/v_{\rm{vir}} = 0.1H{\left( z \right)^{ - 1}}$. The cooling rate can be written as
\begin{equation}\label{eq:coolingrate}
\frac{dm_{\rm{cool}}}{dt}=4\pi\rho_{\rm{g}}(r_{\rm{cool}})r_{\rm{cool}}^2\frac{dr_{\rm{cool}}}{dt}
\end{equation}

As discussed in White \& Frenk (1991), when $r_{\rm{cool}}<R_{\rm{vir}}$, the gas is expected to cool quasistatically from the hot gas
halo. The mass that cools out at each time step may be written as
\begin{equation}\label{eq:coolingmass}
\Delta m_{\rm{cool}}=0.5m_{\rm{hot}}r_{\rm{cool}}v_{\rm{vir}}R_{\rm{vir}}^{-2}\Delta t,
\end{equation}
where $\Delta t$ is the time step. If $r_{\rm{cool}}>R_{\rm{vir}}$, the halo is in the rapid cooling regime, and all the halo gas that
has not already condensed on to galaxies will accrete on to the central object in a so-called ``cold flow''.

Following Mo et al. (1998), the scale length of the disk $r_{\rm{d}}$ that forms through cooling at the centre of a dark matter halo
can be written
\begin{equation}\label{eq:rd}
r_{\rm{d}} = \frac{\lambda}{\sqrt 2 }r_{\rm{vir}}
\end{equation}
in which $\lambda$ is the spin parameter of the dark matter halo, defined as
\begin{equation}\label{eq:lambda}
\lambda=J|E|^{1/2}G^{-1}M_{\rm{vir}}^{-5/2},
\end{equation}
where $J$ and $E$ are the angular momentum and energy of the dark matter component of the halo. The disk dynamical time is given by
\begin{equation}\label{eq:tdyn}
t_{\rm{dyn}}=3r_{\rm{d}}/v_{\rm{vir}}.
\end{equation}

In galaxy disks, cold gas is converted into stars using the formula
\begin{equation}\label{eq:sfr1}
{\dot m_*} =
\begin{cases}
 0 & \left( {{m_{{\rm{gas}}}} < {m_{{\rm{crit}}}}} \right) \\
 \alpha\left(m_{\rm{gas}}-m_{\rm{crit}}\right)/t_{\rm{dyn}} & \left(
{{m_{{\rm{gas}}}} \ge {m_{{\rm{crit}}}}} \right) \\
\end{cases}
\end{equation}
in which the constant $\alpha$ is the star formation efficiency, and $m_{\rm{crit}}$ is the critical mass of gas in the disk.
Following Kennicutt (1989) and Kauffmann (1996), the critical density can be approximately expressed as
\begin{equation}\label{eq:sigmacrit}
\Sigma_{\rm{crit}}=0.59{M_\odot}{\rm{pc}}^{ - 2}\left(\frac{v_{\rm{vir}}}{\rm{km~
s}^{-1}}\right)\left(\frac{r}{\rm{kpc}}\right)^{-1}
\end{equation}
and the critical mass $m_{\rm{crit}}$ (assuming an outer disk radius of $3r_{\rm{d}}$ and a flat gas density profile) is therefore
\begin{equation}\label{eq:mcrit}
m_{\rm{crit}}=5.7\times{10^6}{M_\odot}\left(\frac{v_{\rm{vir}}}{\rm{km~
s}^{-1}}\right)\left(\frac{r_{\rm{d}}}{\rm{kpc}}\right).
\end{equation}
We note that when $m_{\rm{gas}}<m_{\rm{crit}}$, star formation stops over the entire disk. The reason why the star formation rate is
taken to be proportional to $m_{\rm{gas}}-m_{\rm{crit}}$ when $m_{\rm{gas}}>m_{\rm{crit}}$, is to ensure that the star formation rate
is a continuous function of the gas mass.

Energy from supernova explosions can reheat the cold gas in disks to hot gas, and it can also eject some of the hot gas from the halo.
The amount of reheated cold gas is
\begin{equation}\label{eq:mreheated}
\Delta {m_{{\rm{reheat}}}} = {\epsilon_{{\rm{disk}}}}\Delta {m_*},
\end{equation}
where $\Delta {m_*}$ is the mass of stars formed in a given time step, and $\epsilon _{{\rm{disk}}}$ is the supernova reheating
efficiency. The total energy released into feedback processes is approximately
\begin{equation}\label{eq:ereheated}
\Delta {E_{{\rm{SN}}}} = 0.5{\epsilon_{{\rm{halo}}}}\Delta {m_*}v_{{\rm{SN}}}^2,
\end{equation}
where $0.5 v_{{\rm{SN}}}^2$ is the energy in supernova ejecta per unit mass of newly formed stars, and $\epsilon_{{\rm{halo}}}$ is the
efficiency with which the supernovae convert disk gas to halo hot gas. In Equation (\ref{eq:ereheated}), $v_{{\rm{SN}}}=630
\rm{km~s^{-1}}$ is adopted, based on standard supernova theory. If the excess energy in hot gas $\Delta {E_{{\rm{excess}}}}=\Delta
{E_{{\rm{SN}}}}-\Delta {E_{{\rm{reheat}}}}>0$, part of the hot gas will be ejected from the halo. The amount of ejected hot gas is
\begin{equation}\label{eq:ejectedmass}
\Delta m_{\rm{eject}}=\frac{2\Delta E_{\rm{excess}}}{v_{\rm{vir}}^2}=\epsilon_{\rm{halo}}\Delta
{m_*}\frac{v_{\rm{SN}}^2}{v_{\rm{vir}}^2}-\Delta m_{\rm{reheat}}
\end{equation}

The ejected gas will not be lost permanently. As the dark matter halo grows through accretion and merging, the ejected gas will be
reincorporated into the central halo and become part of the hot phase once again. At each time step ($\Delta t$), the amount of
reincorporated gas is
\begin{equation}\label{eq:reincorporatedmas}
\Delta m_{\rm{reincorporate}} = \gamma _{\rm{ej}}(m_{\rm{ejected}}/t_{\rm{dyn}}^{\rm{halo}})\Delta t
\end{equation}
where $m_{\rm{ejected}}$ is the mass of gas currently in the ejecta reservoir and $\gamma_{\rm{ej}}$ is the fraction of the ejected
gas that is reincorporated per halo dynamical time. The default value of $\gamma_{\rm{ej}}$ is 0.5, meaning that half the ejected gas
will be reincorporated into the central halo after one dynamical time (where $t\rm{_{dyn}^{halo}}=R_{vir}/v_{vir}$)

The Instantaneous Recycling Approximation (IRA) is adopted to model the chemical enrichment in the galaxies. Assuming the initial mass
function of Chabrier (2003), the yield is 0.03 and the fraction of the stellar mass returned to the cold interstellar medium is 43\%.
The flow of metals through the cold, hot and ejected components is tracked in exactly the same way as the flow of gas.

In the L-Galaxies code, galaxy mergers are divided into two classes : minor mergers and major mergers. The baryonic mass (gas+star)
ratio of the satellite galaxy to the central galaxy $m_{\rm{sat}}/m_{\rm{cen}}$ determines the merger type. In a minor merger
($m_{\rm{sat}}/m_{\rm{cen}}<0.3$), the stars of the satellite galaxy are added to the bulge of the central galaxy, while the cold gas
of the satellite galaxy is added to the disk of central galaxy. In a major merger ($m_{\rm{sat}}/m_{\rm{cen}}>0.3$), the stellar disks
of both galaxies are destroyed and a bulge is formed. In both minor and major mergers, the mass of stars formed in the associated
starburst is (following Somerville et al. 2001)
\begin{equation}\label{eq:starburst}
\Delta {m_*} = \beta _{\rm{burst}}\left(m_{{\rm{sat}}}/m_{\rm{cen}}\right)^{\alpha_{\rm{burst}}}m_{\rm{gas}}.
\end{equation}
In minor mergers, the stars formed in the burst are added to the disk, but in major mergers all stars formed in the burst are added to
the bulge.

Detailed descriptions of other processes such as reionization, black hole growth and AGN feedback can be found in Croton et al.
(2006), and we inherit these recipes without modification.

\section{Modifications to the Models}

If we wish to study the relations between neutral gas, molecular gas and star formation in galaxies, a single zone treatment of the
gas and stars is not sufficient. In this section, we will describe how we have changed the code so that it is able to track the radial
distribution of gas and stars in galactic disks. Once we are able to specify the surface density of stars and gas as a function of
radius in the disk, it is then a simple matter to incorporate the prescriptions of Blitz \& Rosolowsky (2006) or Krumholz et al.
(2009) to predict the fraction of the gas that is in molecular form.

\subsection{Modelling the radial profiles of the disks} \label{radialprofiles}

\begin{figure*}
\centering
 \includegraphics[angle=-90,scale=0.6]{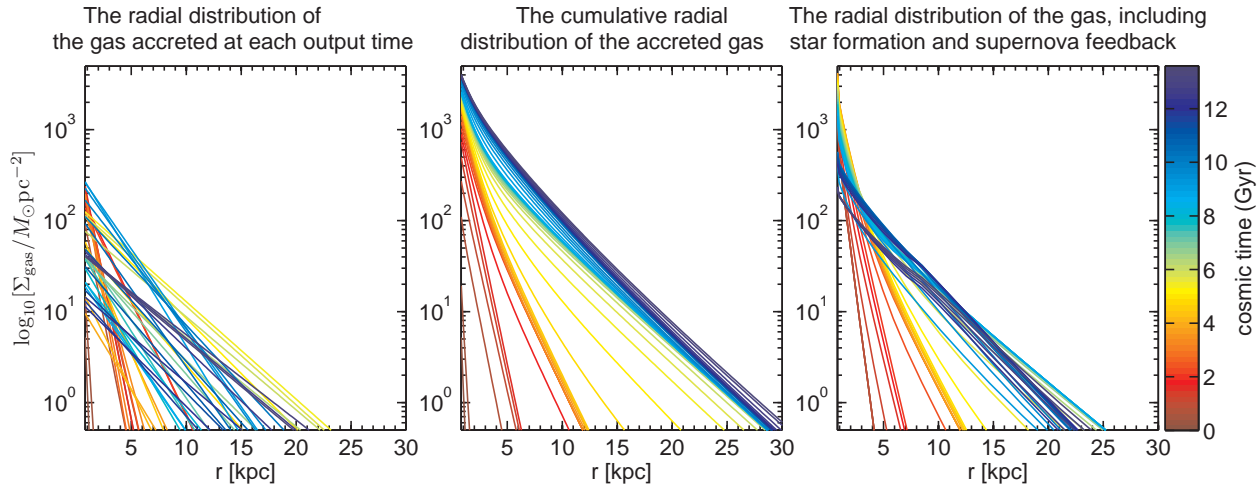}\\
 \caption{Illustration of how our superposition
method works. Results are shown for a ``Milky Way'' type galaxy in a halo with present-day circular velocity $v_{\rm vir} \sim 200$
km/s . In the left panel, we show the radial profiles of the infalling gas for the ``main progenitor'' of the galaxy at each output
time in the simulation. Each profile is colour-coded according to the output time, as indicated in the colour bar on the right. In the
middle panel, we plot the cumulative profile of all the gas that has been accreted up to each output time. In the right panel, we show
the {\em actual} gas profile of the galaxy at each output time, after star formation and supernova feedback processes have been taken
into account.} \label{superposition}
\end{figure*}

In the standard version of the L-Galaxies code, we track the {\em total} stellar mass and gas mass as a function of cosmic time for
each galaxy in the simulation. In our new implementation, we adopt a fixed set of 30 radial ``rings'' for each galaxy, and follow the
build-up of stars and gas within each ring. The radius of each ring is given by the geometric series
\begin{equation}\label{eq:ri}
{r_i} = 0.5 \times {1.2^i}~[h^{-1} \rm{kpc}]~(i=1,2...30)
\end{equation}
where innermost ring has a radius of $0.8$ kpc and the outermost ring has radius $\sim 160$ kpc. We have experimented with other ways
of setting up the rings, for example, arithmetic radial division, and with changing the number of rings. The radial profiles are
insensitive to the precise scheme, so long as the adopted number of rings is sufficiently large.

Following Mo et al. (1998), we assume the gas that cools at a given time step is distributed exponentially with surface density
profile
\begin{equation}\label{eq:exponentialprofile}
{\Sigma _{{\rm{gas}}}}\left( r \right) = {\sgas^0}\exp \left( { - r/{r_{\rm{infall}}}} \right)
\end{equation}
If we assume that the angular momentum of the infalling gas is conserved, the scale length of the infalling gas is $r_{\rm{infall}}=
r_{\rm{d}}=\left(\lambda /\sqrt 2\right)r_{\rm{vir}}$, where $r_{\rm{vir}}$ is the virial radius of the halo and $r_{\rm{d}}$ and
$\lambda$ have the same meaning as in Equations (\ref{eq:rd}) \& (\ref{eq:lambda}). In Equation (\ref{eq:exponentialprofile}),
$\sgas^0$ is the central surface density of the infalling gas and is given by
\begin{equation}\label{eq:sigma0}
{\sgas^0} = \frac{m_{\rm{cool}}}{2\pi r_{\rm{d}}^2}
\end{equation}
where $m_{\rm{cool}}$ is the mass of gas that cools in a given time step. Note that we assume that metals that leave the galaxy disk
are uniformly mixed with the halo gas, which results in a uniform metallicity for the infalling gas.

One important issue that must be addressed is how the gas that accretes at some time $t_i$ is connected to the gas that accretes at a
later time $t_{i+1}$. We adopt the simplest possible superposition scheme, whereby the pre-existing radial profile of the gas remains
unchanged and the new infalling gas is superposed directly on to it.

An illustration of this scheme is presented in Fig. \ref{superposition}. The figure shows the gas accretion history of a ``Milky Way''
type disk galaxy with stellar mass of $M_*=10^{10.7}\ms$ at z=0, residing at the centre of a dark matter halo with virial velocity
$v_{\rm{vir}}=207$km/s. The left panel shows the radial profiles of gas that accretes on to the main progenitor of the galaxy at each
snapshot. The middle panel shows the cumulative profile of all the gas that has been accreted up to that snapshot. The right panel
shows the {\em actual} gas profiles at each snapshot, after star formation and supernova feedback are included. Each curve is
colour-coded according to the time elapsed since the Big Bang: red represents early times and blue represents late times.

At early times, the disk is small and compact. As the Universe evolves, the disks grow in size. The accretion rate of gas on to the
disk reaches a maximum around 4-8 Gyrs after the Big Bang and then declines until the present day. \footnote {This is similar to the
evolution of the global star formation rate density of the Universe (Madau, Pozzetti \& Dickinson 1997), which reaches a maximum
around redshift 0.5-1.5, and then decreases.} In the right panel, we can see how the gas in the central region of the disk is depleted
at late times as the gas is transformed into stars. We note that similar ``inside-out'' disk formation models have been explored in
the past (e.g. Kauffmann 1996; Dalcanton, Spergel \& Summers 1997; Van den Bosch 1998; Avila-Reese, Firmani \& Her{\' a}ndez 1998;
Boissier \& Prantzos 1999; Fu et al. 2009).

When two galaxies merge, we simply add the stellar and gas radial profiles of the satellite galaxy to those of the central galaxy and
then allow for a star burst in each radial bin in the same way as in the original L-Galaxies code. This is probably not an accurate
description of what happens in reality, but our intention in this paper is to concentrate on the predicted gas profiles and scaling
relations of disk-dominated galaxies, which do not experience many mergers.

In this paper, we will not consider the radial distribution of the dark matter. We adopt an approximation of a constant
circular velocity for the whole disk $v_{\rm{cir}}=v_{\rm{vir}}$ when comparing the model results to the observations.

\subsection{The conversion of atomic to molecular gas } \label{h2formation}
Having specified how the infalling gas is distributed across the disk, we are now ready to consider molecular hydrogen formation
processes.

We begin by noting that the total cold gas content of the galaxy predicted by our models includes both hydrogen and helium. The
correction factor between cold hydrogen mass and total cold gas mass is a factor of 1.4 (Arnett 1996). We note that a correction for
helium is applied in some observational analyses, but not in others, so we must be careful to make sure we always compare the models
with the data in a self-consistent way. We also note that recent observations (e.g Reynolds 2004) of the Local Interstellar Clouds in
the Milk Way disk show that about 1/3 of all the gas in the Milky Way disk is in a warm, ionized phase. Following Obreschkow et al.
(2009), we include a warm phase correction into our models. The warm phase coefficient is $\xi=1.3$. If there is no note to the
contrary, the gas surface densities are always divided by $\xi$ in this paper.

We have implemented two different molecular gas formation prescriptions in our models. The first prescription is from Krumholz et al.
(2009) (hereafter K09), in which the $\h2$ fraction is a function of local cold gas surface density and metallicity (hereafter $\h2$
prescription 1). The other prescription has its origin in papers by Elmegreen (1989 \& 1993), Blitz \& Rosolowsky (2006), and
Obreschkow et al. (2009), which have all suggested that the $\h2$ fraction is a function of the pressure of the ISM (hereafter $\h2$
prescription 2).

\subsubsection{The $\h2$ prescription of Krumholz et al.} \label{h2fraction1}

Krumholz et al. (2008) considered the location of the atomic-to-molecular transition in a uniform spherical gas cloud
bathed in a uniform, isotropic, dissociating radiation field. The main result of their calculation is that the amount
of atomic material required to shield a molecular cloud against dissociation by the interstellar radiation field is
characterized by two parameters: $\tau_R$, a measure of the dust optical depth of the cloud, and $\chi$, a normalized
radiation field strength, which is defined as the ratio of the rate at which Lyman-Werner photons are absorbed by dust
grains to the rate at which they are absorbed by hydrogen molecules in a parcel of predominantly atomic gas in
dissociation equilibrium in free space.

In Krumholz et al. (2009) and McKee \& Krumholz (2010), this idealized model was applied to atomic-molecular complexes in galaxies in
an attempt to elucidate the physical processes and parameters that determine the molecular content in these systems. The paper
demonstrates that because of the way the density in the cold phase of the atomic ISM varies with the interstellar radiation field,
$\chi \sim 1$ in all galaxies with only a weak dependence on metallicity. The existence of a ``characteristic'' normalized radiation
field strength, leads to a simple analytic approximation for the fraction of mass in an atomic-molecular complex that is in the
molecular phase solely in terms of the column density of the complex and the metallicity of the gas. Krumholz et al. (2009) show that
the predictions of their models agree well with observations of the atomic and molecular content of the cold gas in the Milky Way and
in nearby galaxies, particularly in regions where $\sgas > 10\mspc$.

In our model, we adopt the molecular fraction values $\fh2(\sgas,\zgas)$ from K09 to calculate the mass of neutral and molecular gas
in each ring, so that
\begin{equation}\label{eq:h2h1fraction}
\begin{array}{l}
 {\Sigma_{{{\rm{H}}_2}}} = {f_{{{\rm{H}}_2}}}{\Sigma_{{\rm{gas}}}} \\
 {\Sigma_{{\rm{HI}}}} = \left( {1 - {f_{{{\rm{H}}_2}}}} \right){\Sigma_{{\rm{gas}}}} \\
 \end{array}
\end{equation}

Since the gas surface density defined in K09 is a {\em local} surface density, while $\sgas$ in our models is an azimuthally-averaged
surface density, we introduce a clumping factor $c_{\rm{f}}$ to account for the fact that the gas in disks is usually organized into
higher density clumps in spiral structures. In this case, $\fh2$ should be written as
\begin{equation}\label{eq:h2fraction1}
\fh2=\fh2(c_{\rm{f}}\sgas,\zgas).
\end{equation}
In our models, $c_f$ is left as a free parameter (see section 3.5 for a description for how the free parameters are set).

\subsubsection{The pressure related $\h2$ prescription} \label{h2fraction2}

Blitz \& Rosolowsky (2004 \& 2006) proposed that pressure alone determines the ratio of atomic to molecular gas averaged over a
particular radius in disk galaxies. In these papers, the molecular ratio of a galaxy disk was expressed as
\begin{equation}\label{eq:BRh2}
R_{\rm{mol}}(r) \equiv {M_{{{\rm{H}}_2}}}\left( r \right)/{M_{{\rm{HI}}}}\left( r \right) = {\left[ {P\left( r
\right)/{P_0}} \right]^{\alpha_P} },
\end{equation}
where $P_0$ and $\alpha_P$ are constants fit from the observations.

We follow a similar procedure that outlined in Obreschkow et al. (2009) to calculate the pressure in our model disk galaxies.
According to Elmegreen (1989 \& 1993), the midplane-pressure of the ISM in disk galaxies can be expressed as
\begin{equation}\label{eq:elmegreenpressure}
P\left( r \right) = \frac{\pi }{2}G{\Sigma _{{\rm{gas}}}}\left( r \right)\left[ {{\Sigma _{{\rm{gas}}}}\left( r \right)
+ {f_\sigma }\left( r \right){\Sigma _*}\left( r \right)} \right],
\end{equation}
where $G$ is the gravitational constant, $r$ is the radius from the galaxy centre, and ${f_\sigma}(r)$ is the ratio of the vertical
velocity dispersions of the gas and the stars:
\begin{equation}\label{eq:fsigma}
{f_\sigma }\left( r \right) = {\sigma_{{\rm{gas}}}}\left( r \right)/{\sigma _*}\left( r \right).
\end{equation}

Observations indicate that $\sigma _{\rm{gas}}$ is approximately a constant across the whole disk (e.g. Boulanger \& Viallefond 1992;
Leroy et al. 2008). $\sigma _*(r)$ decreases exponentially with a scale length twice that of the stellar disk (e.g Bottema 1993), i.e

\begin{equation}\label{eq:sigmastar}
\sigma_*(r)=\sigma{_*^0}\exp \left( -r/2r_* \right) ,
\end{equation}
where $\sigma{_*^0}$ is the stellar velocity dispersion at the centre of the disk. Substituting the exponential stellar disk profile
${\Sigma _*}\left( r \right) = \Sigma{_*^0}\exp \left( { - r/{r_*}} \right)$ and Equation (\ref{eq:sigmastar}) into Equation
(\ref{eq:fsigma}), we get
\begin{equation}\label{eq:fsigma1}
f_{\sigma}(r) = \frac{\sigma _{\rm{gas}}}{\sigma{_*^0}}\sqrt {\frac{\Sigma{_*^0}}{\Sigma _*(r)}} \equiv {f_{\sigma
}^0}\sqrt {\frac{\Sigma{_*^0}}{{\Sigma _*}(r)}}
\end{equation}
where $f_{\sigma}^0$ is the value of $f_{\sigma}(r)$ at the centre of the disk. The mean value of $f_{\sigma}(r)$ for the whole disk
is
\begin{equation}\label{eq:fsigmamean}
{\bar f_\sigma } = \frac{{\int {2\pi r{\Sigma _*}\left( r \right){f_\sigma }\left( r \right)dr} }}{{\int {2\pi r{\Sigma
_*}\left( r \right)dr} }} = \frac{{{\Sigma _*^0}{f_{\sigma}^0}{{\left( {2{r_*}} \right)}^2}}}{{{\Sigma _*^0}r_*^2}} =
4f_{\sigma}^0
\end{equation}
Adopting the value quoted by Elmegreen (1993), ${\bar f_\sigma } \approx 0.4$, we get $f_{\sigma}(r) \approx 0.1\sqrt
{\Sigma_*^0/{\Sigma _*}(r)}$. Substituting into Equation (\ref{eq:elmegreenpressure}), the pressure of ISM may be written
\begin{equation}\label{eq:pressure}
P(r) = \frac{\pi }{2}G{\Sigma _{{\rm{gas}}}}\left( r \right)\left[ {{\Sigma _{{\rm{gas}}}}\left( r \right) + 0.1\sqrt
{{\Sigma _*}\left( r \right){\Sigma_*^0}} } \right]
\end{equation}

Equation (\ref{eq:pressure}) can then be substituted into Equation (\ref{eq:BRh2}) to obtain the molecular ratio. We adopt ${P_0} =
5.93 \times {10^{ - 13}}$Pa and $\alpha_P=0.92$ from Blitz \& Rosolowsky (2006), which were the mean values obtained from fitting
Equation (\ref{eq:BRh2}) to a small sample of nearby galaxies. Our final expression for the molecular fraction is
\begin{multline}\label{eq:rmol}
{f_{{\rm{H_2}}}}\left( r \right)=\\
1.38 \times {10^{-3}}{\left[ {\Sigma_{\rm{gas}}^2(r)+0.1{\Sigma _{\rm{gas}}}(r)\sqrt{\Sigma _*(r)\Sigma _*^0} } \right]^{0.92}},
\end{multline}
where $\sgas(r)$, $\Sigma_*(r)$ are the radial gas and stellar surface densities, $\Sigma_*^0$ is the central stellar surface density
(units are $\mspc$). Thus, in the pressure-based prescription, the molecular ratio depends primarily on the gas surface density, but
the stellar surface density can be important, particularly in gas-poor disk galaxies at low redshifts.

Finally we note that when we apply our pressure related prescription to the disks in our simulation, we do not apply any clumping
factor or warm phase corrections, because Equation (\ref{eq:BRh2}) refers to the total pressure averaged over a particular
galactocentric radius.

\subsubsection{Comparison between the two $\h2$ fraction prescriptions} \label{h2fractioncomp}

\begin{figure*}
\centering
 \includegraphics[angle=-90,scale=0.55]{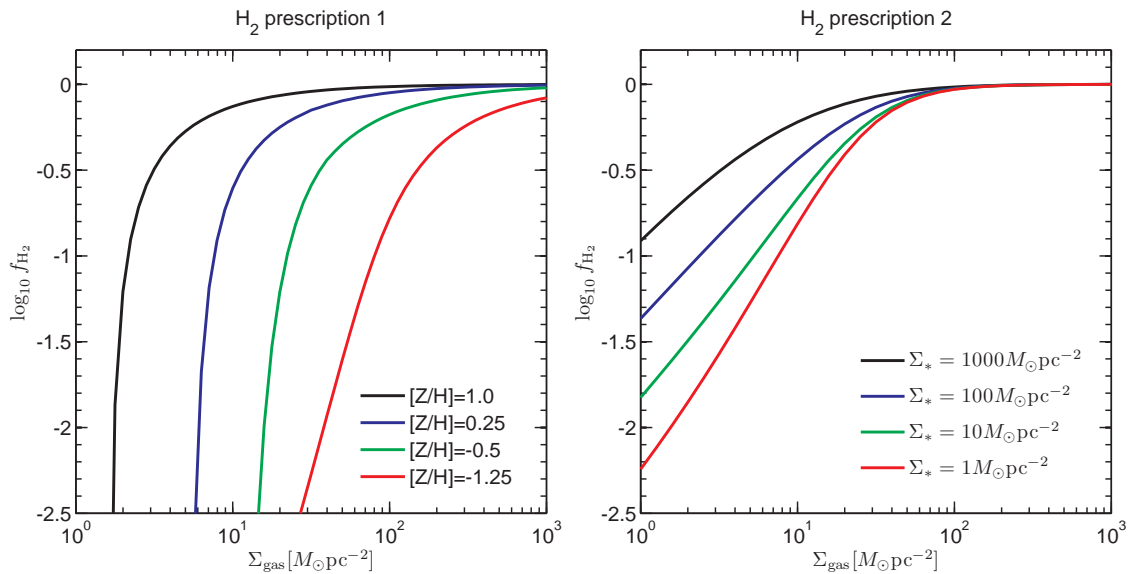}\\
 \caption{ Comparison of the results for the two $\h2$ prescriptions.
The left panel shows the relation between $\h2$ fractions and local gas surface density predicted by the Krumholz et al. models for
four different gas metallicities. The right panel shows the relation between $\h2$ fraction and local gas surface density predicted by
the pressure-based prescription for four different values of the stellar surface density. A central stellar surface density
$\Sigma_*^0=2000 \mspc$ is assumed in the right panel.
 }\label{h2fractioncompare}
\end{figure*}

Fig. \ref{h2fractioncompare} illustrates the main differences between the two $\h2$ prescriptions described above. On the left, we
plot $\fh2$ vs. $\sgas$ for the Krumholz prescription; each curve corresponds to a different value of the gas-phase metallicity. On
the right, we plot $\fh2$ vs. $\sgas$ for the pressure-based prescription; each curve corresponds to a different stellar surface
density (note that we assume a central stellar surface density of $\Sigma_*^0=2000\mspc$ in Equation (\ref{eq:rmol}), which is typical
for a spiral galaxy like the Milky Way).

We see that the molecular fraction decreases at low gas surface densities in both panels, but the thresholding effect is much stronger
for the Krumholz prescription than for the pressure-based prescription. In addition, the Krumholz prescription appears to open up the
possibility that molecular gas formation is strongly suppressed in low mass galaxies, which have both low densities and low
metallicities. We will come back to these points in future work.

\subsection{Star formation}

In a recent paper, Leroy et al. (2008) measured the local star formation efficiency (SFE) (i.e the star formation rate (SFR) per unit
mass in gas) for a small sample of nearby galaxies and compared it with expectations from a number of proposed star formation laws.
Their basic result is that in the inner regions of spirals, where the molecular gas constitutes a significant fraction of the total
cold gas content of the galaxy, the SFE of $\h2$ is nearly constant at $(5.25 \pm 2.5) \times 10^{-10}$ yr$^{-1}$ (corresponding to a
$\h2$ depletion time about $2 \times 10^9$ yr). Leroy et al. tested whether there were additional dependencies on variables such as
free fall and orbital time-scales, midplane gas pressure, stability of the gas disk to collapse, and the ability of perturbations to
grow despite shear, but they found no effect.

These results lead us to implement the following star formation ``law'' in our models:
\begin{equation}\label{eq:h2sfr}
\Sigma_{\rm{SFR}}=\alpha\Sigma_{\rm{H_2}},
\end{equation}
where the molecular star efficiency $\alpha$ is a constant for all galaxies, and we adopt $\alpha=4.9\times{10^{- 10}}{\rm{yr}}^{-1}$,
similar to the value in Leroy et al. (2008).

\subsection{Constraints from the radial profiles} \label{problem}

\begin{figure}
  \includegraphics[angle=0,scale=0.51]{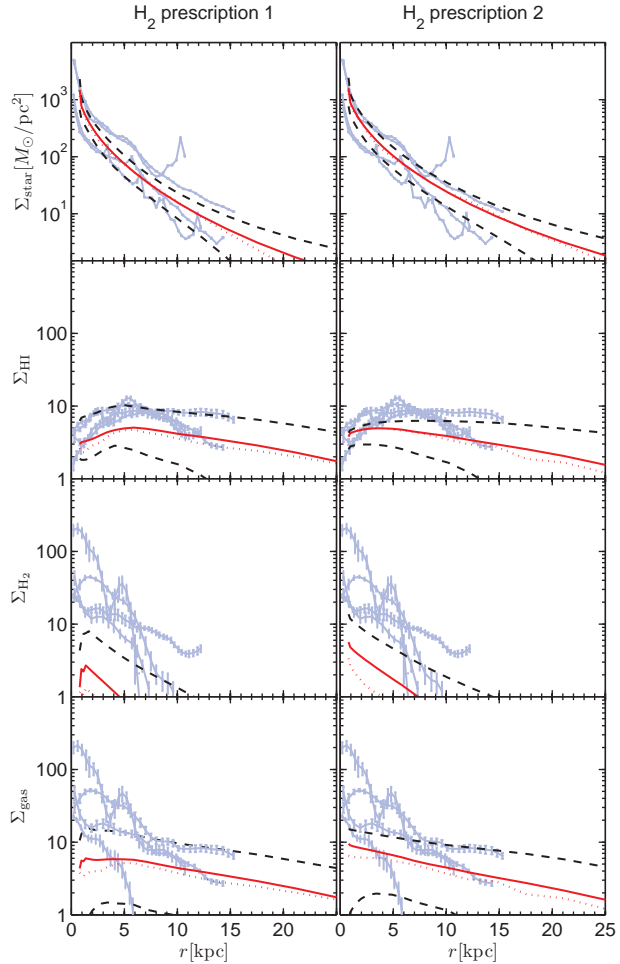}
  \caption{The radial surface density profiles of stars, cold gas, HI and $\h2$ for galaxies
  with masses similar to that of the Milky Way at z=0. The light blue curves with error bars are taken from the data
  presented in Leroy et al. (2008) and include galaxies with circular velocities in the range
  200km/s$<v_{\rm{vir}}<$235 km/s and $M_*^{\rm{bulge}}/M_*\le 15\%$. The red solid and dotted curves are the mean
  and median values from the models, and the black dashed curves show the $\pm1\sigma$ deviations about the mean.
  The panels on the left show the model results when $\h2$ prescription 1 is used, and the panels on the right are
  for $\h2$ prescription 2.
  }\label{h2profiles}
\end{figure}

In Fig. \ref{h2profiles}, we plot the radial surface density profiles of stars, HI, $\h2$ and total cold gas for ``Milky Way'' sized
disk galaxies residing in dark matter haloes with circular velocities in the range 200--235km/s at redshift 0. We show results for
both $\h2$ fraction recipes and compare with data from Leroy et al. (2008). We find that we are unable to get a good fit to the data,
even if we adjust all available model parameters. Although we can get stellar surface density profiles that match the observations
very well, the HI and total gas surface density profiles are too flat. In addition, the $\h2$ surface density is more than an order of
magnitude too low. The main reason for this discrepancy is that a star formation law of the form given in equation (\ref{eq:h2sfr})
leads to a gas consumption rate that is {\em steeper} than the total gas profile, resulting in rapid depletion of the gas in the inner
part of the disk.

Recall that the amount of gas in the disk is regulated not only by star formation, but also by supernova feedback processes, which act
to heat the cold gas. The supernova reheating rate is proportional to the mass of newly formed stars (Equation \ref{eq:mreheated}), so
Equation (\ref{eq:h2sfr}) then implies that the net gas consumption rate will proportional to the surface density of $\h2$. Because
the $\h2$ fraction is a strongly increasing function of gas surface density, the $\h2$ surface density profiles are always steeper
than the total gas surface density profiles. As a result, the gas consumption rate has a steeper dependence on radius than the gas
profile itself, and this causes the gas profiles to flatten with time.

One possible solution to this problem is that the gas does not remain fixed at a given radius, but flows inwards, thus replenishing
the gas that is consumed at the centre of the galaxy by star formation. We do not consider radial gas flows in this paper. Instead, we
make two plausible changes to the star formation ``law'' and feedback prescriptions that act to bring the gas profiles in better
agreement with observations.

First, we are computing $\fh2$ based on azimuthally-averaged column densities, but this approximation becomes increasingly poor in the
outer parts of galaxies where the mean column density and mean molecular fraction are very low. In such regions the star formation
tends to be dominated by small isolated regions where the {\em local} column density is much higher than the azimuthally-averaged
value, and as a result the molecular fraction is higher. Since we cannot easily capture this effect with our azimuthally-averaged
prescription, we instead modify the star formation law based on an empirical fit to the behavior of the star formation rate surface
density in the outer regions of galactic disks (Bigiel et al 2008). We adopt
\begin{equation}\label{eq:sfr}
\Sigma_{\rm{SFR}} =
\begin{cases}
   \alpha\Sigma_{\rm{H_2}} & (f_{\rm{H_2}}\ge 0.5)  \\
   \alpha'\Sigma_{\rm{gas}}^2 & (f_{\rm{H_2}} < 0.5),  \\
\end{cases}
\end{equation}
where $\alpha$ has the same value as in Equation (\ref{eq:h2sfr}), and $\alpha'=0.5\alpha/\Sigma_{\rm{gas}}|_{\fh2=0.5}$, so that the
star formation rate changes continuously at the radius where $\fh2=0.5$. In other words, our proposed ``hybrid'' star formation law
results in a star formation rate that is proportional to the $\Sh2$ surface density in the regions of the disk where $\h2$ is
dominant, but a Schmidt-Kennicutt type star formation law ${\Sigma_{\rm{SFR}}}\propto\sgas^n$ with $n=2$ is adopted in the HI dominant
regions.

\begin{figure}
\centering
  \includegraphics[angle=-90,scale=0.6]{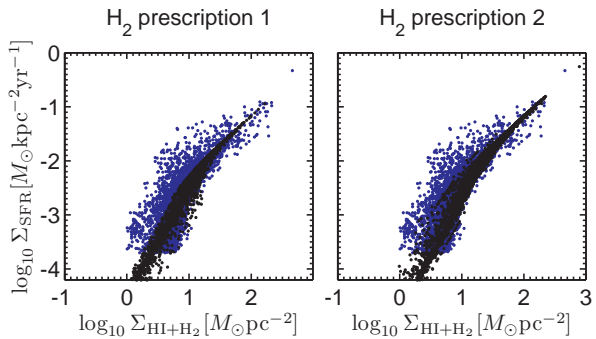}
  \caption{Black points show the star formation rate surface density as a function of the HI+$\h2$ surface density
   within each radial ring for a set of randomly selected galaxies from our simulation at z=0. Blue points are
   observational data taken from Figure 10 of Bigiel et al. (2008). Results are shown for both $\h2$ fraction prescriptions.
  }\label{sfr}
\end{figure}

In Fig. \ref{sfr}, we plot $\Sigma_{\rm{SFR}}$ versus $\Sigma_{\rm{HI}+\h2}$ for a set of randomly selected galaxies from our
simulation at z=0. The black points represent gas surface densities and star formation rates evaluated in the 30 radial rings in each
of the selected galaxies. Results are shown for both $\h2$ fraction prescriptions. In order to be consistent with the data in Bigiel
et al. (2008), the helium component is not included in Fig. \ref{sfr}. For comparison, we plot the observed relation between
$\Sigma_{\rm{SFR}}$ and $\Sigma_{\rm{HI}+\h2}$ from Bigiel et al. (2008). As can be seen, our ``hybrid'' SFR law provides an excellent
match to the real data over the entire range is total gas surface density probed by the observations. A change in slope occurs at
$\Sigma_{\rm{HI}+\h2}\approx 10 \mspc$, which marks the division between the $\h2$ and HI dominant regions of the galaxy.

The change in star formation law helps to steepen the gas profiles in the outer disk, but does not solve the problem of the
over-consumption of gas in the central regions of the galaxy. In our standard supernova feedback recipe, the mass of cold gas reheated
by supernovae is proportional to the mass of newly formed stars. We might hypothesize that the dissipation of energy input by
supernovae is more efficient in denser regions (Equation \ref{eq:coolingtime}), so that less gas is reheated in the inner regions of
the disk per unit mass of stars that are formed. We have chosen to make a correction to Equation (\ref{eq:mreheated}) so that the
reheated mass is inversely proportional to the surface density of gas,
\begin{equation}\label{eq:mreheatedgas}
\Delta m_{\rm{reheat}}=\epsilon_{\rm{disk}} \frac{\Sigma_{\rm{0gas}}}{\Sigma_{\rm{gas}}}\Delta m_*,
\end{equation}
where the coefficient $\Sigma_{\rm{0gas}}$ is the same for all galaxies.

In the remainder of this paper, we will always adopt Equations (\ref{eq:sfr}) \& (\ref{eq:mreheatedgas}) to describe star formation
and supernova reheating, respectively.

\subsection{Model parameters}

\begin{table*}
 \centering
 \caption{The model parameters \label{tab:model parameters}}
 \begin{tabular}{|c|l|l|}
 \hline \hline
Parameter & Description & Value\\
 \hline
$c_{\rm{f}}$ & clumping factor for $\h2$ prescription 1 (Eq. \ref{eq:h2fraction1}) & 1.5 \\
$\xi$ & warm phase correction factor (Sec. \ref{h2formation})& 1.3 \\
$P_0$ & the constant of the relation between molecular ratio and ISM pressure (Eq. \ref{eq:BRh2})& $5.93\times10^{-13}$Pa \\
$\alpha_P$ & the index of the relation between molecular ratio and ISM pressure (Eq. \ref{eq:BRh2}) & 0.92\\
$\alpha$ & star formation efficiency in (Eq. \ref{eq:h2sfr} \& \ref{eq:sfr}) & $4.9\times{10^{- 10}}{\rm{yr}}^{-1}$ \\
$\epsilon_{\rm{disk}}$ & the supernova reheating rate (Eq. \ref{eq:mreheated} \& \ref{eq:mreheatedgas}) & 3.5 \\
$\Sigma_{\rm{0gas}}$ & constant surface density in the supernova reheating recipe (Eq. \ref{eq:mreheatedgas}) & $2.1\mspc$ \\
$\kappa_{\rm{AGN}}$ & Quiescent hot gas black hole accretion rate & $2.0\times{10^{-6}}\ms\rm{yr^{-1}}$\\
\hline \hline
\end{tabular}
\end{table*}

In Table \ref{tab:model parameters}, we list the values of the new model parameters that we introduced: $c_{\rm{f}}$, $\xi$, $P_0$,
$\alpha_P$, $\alpha$, $\epsilon_{\rm{disk}}$, $\Sigma_{\rm{0gas}}$ and $\kappa_{\rm{AGN}}$. The values of the other parameters in the
L-Galaxies code are unchanged and they can be found in Table 1 in Croton et al. (2006). Note that the values of $\xi$, $P_0$ and
$\alpha_P$ are adopted directly from observations and are not tunable parameters. $c_{\rm{f}}$, $\alpha$, $\epsilon_{\rm{disk}}$,
$\Sigma_{\rm{0gas}}$ and $\kappa_{\rm{AGN}}$ are free parameters in our models, which we adjust to fit both the observed stellar and
gas density profiles and the global stellar and gas mass functions (see Sec. 4.2). We will now clarify how each of these parameters
affects our main results:

$c_{\rm{f}}$: As discussed in Sec. \ref{h2fraction1}, the clumping factor corrects for the difference between the local gas surface
density that is relevant to the Krumholz et al. $\h2$ fraction prescription, and the azimuthally-averaged gas surface densities
predicted by the models. Higher values of $c_{\rm{f}}$ lead to higher predicted $\h2$ fractions. The left panel of Fig.
\ref{h2fractioncompare} shows that a change in $c_{\rm f}$ has the greatest effect in the low gas surface density regions of the disk.
In practice, we tuned $c_{\rm{f}}$ so as to best reproduce the observed $\h2$ and HI mass functions at $z=0$. We found that $c_{\rm
f}=1.5$ allowed us to fit both the gas surface density profiles of Milky Way-type galaxies and the gas mass functions at z=0.

$\alpha$: Leroy et al. (2008) find an $\h2$ star formation efficiency $\alpha=(5.25 \pm 2.5)\times 10^{-10}$ yr$^{-1}$ from their
observational data. In our models, the value of $\alpha$ controls the amplitude of the stellar mass profiles and the stellar mass
function. Our adopted value of $\alpha=4.9\times 10^{-10}$ yr$^{-1}$ allows us to fit the stellar mass function at $z=0$ and is in
remarkably good agreement with the value found by Leroy et al (2008).

$\epsilon_{\rm{disk}}$ and $\Sigma_{\rm{0gas}}$: From Equation (\ref{eq:mreheatedgas}), we see that
$\epsilon_{\rm{disk}}\Sigma_{\rm{0gas}}$ controls the mass of cold gas reheated by supernova explosion. Higher values of
$\epsilon_{\rm{disk}}\Sigma_{\rm{0gas}}$ lead to lower gas surface densities and a lower amplitude of the gas mass functions. We have
chosen to fix $\epsilon_{\rm{disk}}$ at a value of 3.5 (the same as in Croton et al. 2006). We tune $\Sigma_{\rm{0gas}}$ to fit the
gas surface density profiles and gas mass functions.

$\kappa_{\rm{AGN}}$: is the quiescent hot gas black hole accretion rate, which controls the fraction of hot gas in massive dark matter
haloes that condense on to central galaxy that host super massive black holes. Higher values of $\kappa_{\rm{AGN}}$ decrease the
accretion rate of gas from the surrounding halo and suppress the number of very high mass galaxies. Because we changed the way star
formation is treated in the model, we retuned $\kappa_{\rm{AGN}}$ to obtain good fit to the stellar mass function at the high mass
end.

\section{Results}

\subsection{Radial surface density profiles at redshift=0} \label{radialsurfacedensity}

\begin{figure}
  \includegraphics[angle=0,scale=0.53]{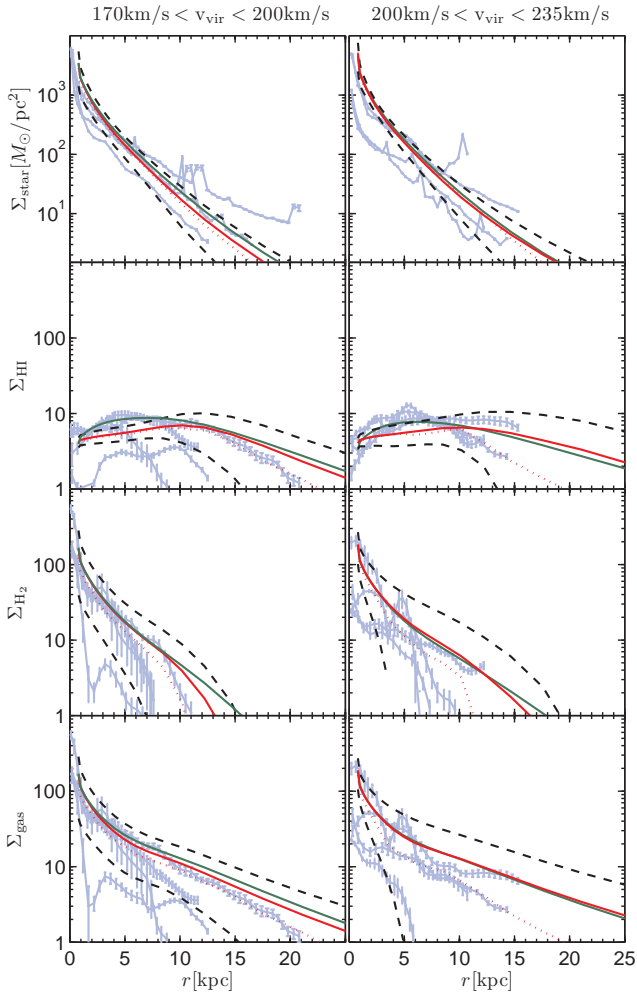}
  \caption{The radial surface density profiles of stars, HI, $\h2$, and total cold gas for disk galaxies in our models
  at redshift 0. The light blue curves with error bars are from the observational data of Leroy et al. (2008).
  The red curve shows the mean profile for $\h2$ prescription 1, while the dashed black curves show the
  $\pm 1 \sigma$ deviations about the mean. The green curves show the mean profiles for $\h2$ prescription 2. All
  the gas surface density profiles include the contribution from helium and a correction for the warm phase. The
  right column shows results for disk galaxies with rotation velocities similar to the Milky Way
  (200 km/s$<v_{\rm{vir}}<$235 km/s), and the left column for lower mass disk galaxies (170 km/s$<v_{\rm{vir}}<$200 km/s).
}\label{model compare}
\end{figure}

In Fig. \ref{model compare}, we compare the radial surface density profiles of stars, HI, $\h2$ and the total cold gas
in disk galaxies at redshift z=0 with the observational results of Leroy et al. (2008). In the right panels, we show
results for galaxies with circular velocities similar to the Milky Way (200 km/s$<v_{\rm{cir}}<$235 km/s), and in the
left panel we show results for disk galaxies with somewhat lower circular velocities (170 km/s$<v_{\rm{cir}}<$200
km/s). As mentioned in Sec. \ref{radialprofiles}, we assume constant disk radial velocity profile
$v_{\rm{cir}}=v_{\rm{vir}}$ for simplicity. The Leroy et al. (2008) data compilation consists of $\Sigma_*$ profiles
derived from data from the Spitzer Infrared Nearby Galaxies Survey (SINGS) (Kennicutt et al. 2003), $\SHI$ profiles
from The HI Nearby Galaxy Survey (THINGS) (Walter et al. 2003), and $\Sh2$ profiles from the HERACLES (Leroy et al.
2008) and BIMA SONG (Helfer et al. 2003) surveys. Four of the galaxies in the Leroy et al. (2008) sample have circular
velocities in the range 200 km/s$<v_{\rm{cir}}<$235 km/s: NGC 0628, NGC 3184, NGC 5194, and NGC 3521; another four have
170km/s$<v_{\rm{cir}}<$200km/s: NGC 3351, NGC 6946, NGC 3627, and NGC 5055. All 8 galaxies are spiral galaxies with
morphological type later than Sb (Hubble type index T $\ge 3$). Laurikainen et al. (2007) and Weinzirl et al. (2009)
have quantified the average relation between Hubble type and the ratio of the luminosities of bulges to galaxy disks.
The fraction of the mass of the galaxy in the bulge ($B/T \equiv M_*^{\rm{bulge}}/M_*$) for spiral galaxies later than
Sb is typically lower than 0.15. Thus, we select the model galaxies in dark matter haloes with virial velocities in the
range 200 km/s$<v_{\rm{vir}}<$235 km/s and 170 km/s$<v_{\rm{vir}}<$200 km/s with $B/T\le 15\%$ to compare to the
observations. In each sub-panel of Fig. \ref{model compare}, the mean profiles for the models that assume the Krumholz
$\h2$ fraction prescription are plotted as red solid curves, the median profiles as a red dotted curves, and the
$\pm1\sigma$ deviations about the mean as black dashed curves. The mean profiles for the models that assume the
pressure-based $\h2$ fraction prescription are plotted as green solid curves.

In Fig. \ref{model compare}, we see that the mean surface density profiles of the gas and the molecular gas in the inner regions of
the disk are reasonably well fit by an ``exponential law''. This is mostly a consequence of the fact that the infalling gas is assumed
to have an exponential profile at each time step. Fig. \ref{model compare} shows that the problem with the overconsumption of
molecular gas in the inner disk discussed in Section \ref{problem} has been solved by the new implementation of supernova feedback
discussed in the previous section. In agreement with observations, the molecular gas dominates in the inner disk, the neutral gas
dominates in the outer disk, and the $\h2$ profiles closely track the stellar profiles. Comparing the red and green curves, we see
that the two $\h2$ fraction prescriptions give very similar results. Note that we have tuned the available free parameters to
reproduce the observational data as well as possible, so this is not entirely surprising. The main differences between the two
prescriptions occur in the outer disks, where $\Sh2$ falls off less steeply for the pressure-based prescription, particularly for
galaxies with lower circular velocities, which have lower metallicities. The results for inner disks are almost identical for the two
prescriptions.

\subsection{Stellar and gas mass functions at z=0} \label{massfunctions}

\begin{figure*}
  \centering
  \includegraphics[angle=-90,scale=0.62]{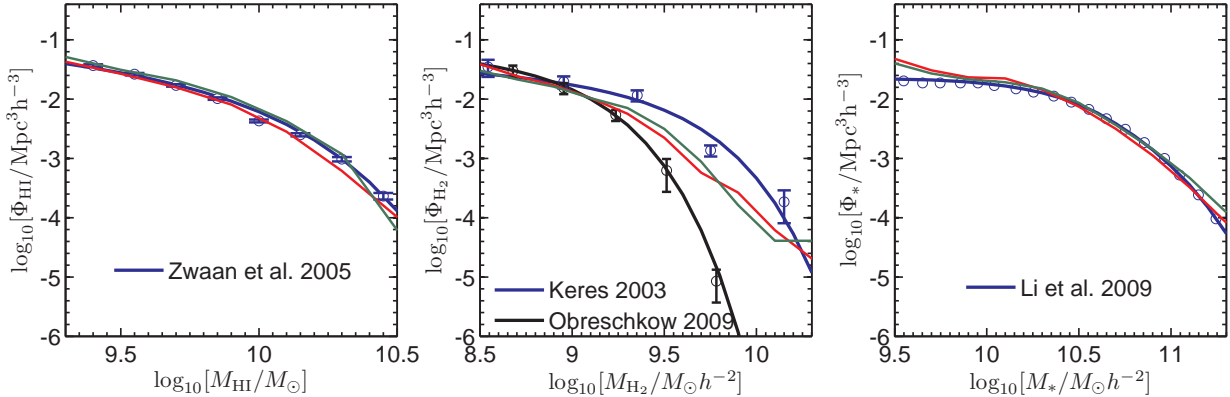}\\
  \caption{The mass functions of HI, $\h2$ and stars for galaxies at redshift zero. The red and green curves are model
  results for $\h2$ prescription 1 and 2, respectively. The HI and $\h2$ mass functions include the correction for the warm phase
  but do not include the helium component. The open symbols with error bars indicate the observational data, while the blue and
  black curves indicate the best-fit Schechter functions. The stellar mass function is taken from Li et al. (2009)
  analysis of the final data from the Sloan Digital Sky Survey. The HI mass function was derived from HIPASS by Zwaan
  et al. (2005). The $\h2$ mass function is from the FCRAO Extragalactic CO Survey; the blue circles and curves assume
  a fixed CO-$\h2$ conversion factor (Keres, Yun \& Young 2003), while the black curves assume a variable CO-$\h2$ conversion
  factor as given in Obreschkow \& Rawlings (2009).
  }\label{mass function}
\end{figure*}

In Fig. \ref{mass function}, we show the HI, $\h2$ and the stellar mass functions predicted by our models at $z=0$. The results are
compared to the HI mass function derived from the HI Parkes All Sky Survey (HIPASS) by Zwaan et al. (2005), the stellar mass function
derived from the data release 7 of the SDSS by Li et al. (2009), and the $\h2$ mass function derived from the Five College Radio
Astronomy Observatory (FCRAO) Extragalactic CO Survey by Keres et al. (2003) assuming a constant CO-$\h2$ conversion factor (blue),
and by Obreschkow \& Rawlings (2009) assuming a variable CO-$\h2$ conversion factor that scales with the B-band magnitude of the
galaxy, as proposed by Boselli, Lequeux \& Gavazzi (2002) (black). The red curves in Fig. \ref{mass function} are our model results
with $\h2$ prescription 1, while the green curves show the results for $\h2$ prescription 2.

Note that we only plot the stellar mass function down to a limiting mass of $M_*=10^{9.5}\ms$, for which the Millennium Simulation has
sufficient resolution to accurately track the formation history of the host halo. The stellar mass functions predicted by the models
agree very well with the data for galaxies with stellar masses greater than $10^{10} M_{\odot}$. The models produce too many galaxies
with masses less than $10^{10} M_{\odot}$. Guo et al. (2010) have implemented the galaxy formation model of De Lucia \& Blaizot (2007)
in a higher resolution simulation, and show that the mismatch with the faint end of the galaxy mass function becomes increasingly
severe as one goes to lower masses. Since we are not concerned with dwarf galaxies in this paper, we will ignore this problem for the
moment.

The agreement with the HI mass function of Zwaan et al. (2005) is excellent. It is also interesting that the $\h2$ mass function
predicted by the model is in closer agreement with the observational results derived using a constant conversion factor.

Finally, we note that the stellar and cold gas mass functions predicted by the models at z=0 are insensitive to the choice of $\h2$
fraction prescriptions.

\section{Global scaling relations between the molecular gas, neutral gas and stellar masses of galaxies}

In this section, we investigate the correlations between the atomic gas fraction $\HIs$, the molecular gas fraction $\Hs$ and the
molecular-to-atomic gas ratio $\HHI$ with galaxy properties such as stellar mass, average stellar surface density and average gas
surface density. We then elucidate the physical processes that determine the slope and scatter of these relations. Finally, we compare
the predictions of the models with results derived from recent data sets.

\subsection{Model results}

\begin{figure*}
\flushleft
  \subfigure[]{
  \includegraphics[angle=0,scale=0.52]{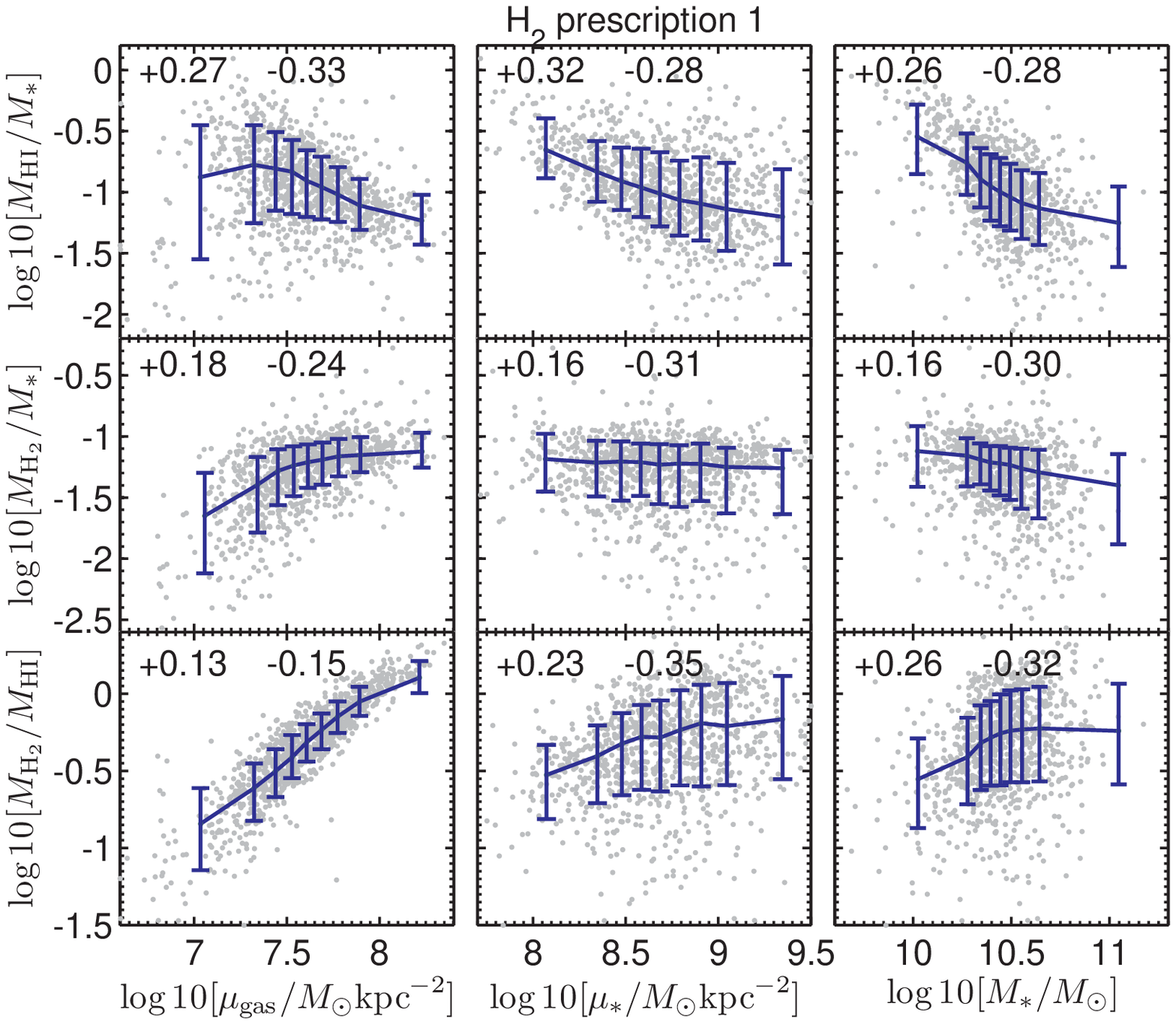}}
  \subfigure[]{
  \includegraphics[angle=0,scale=0.52]{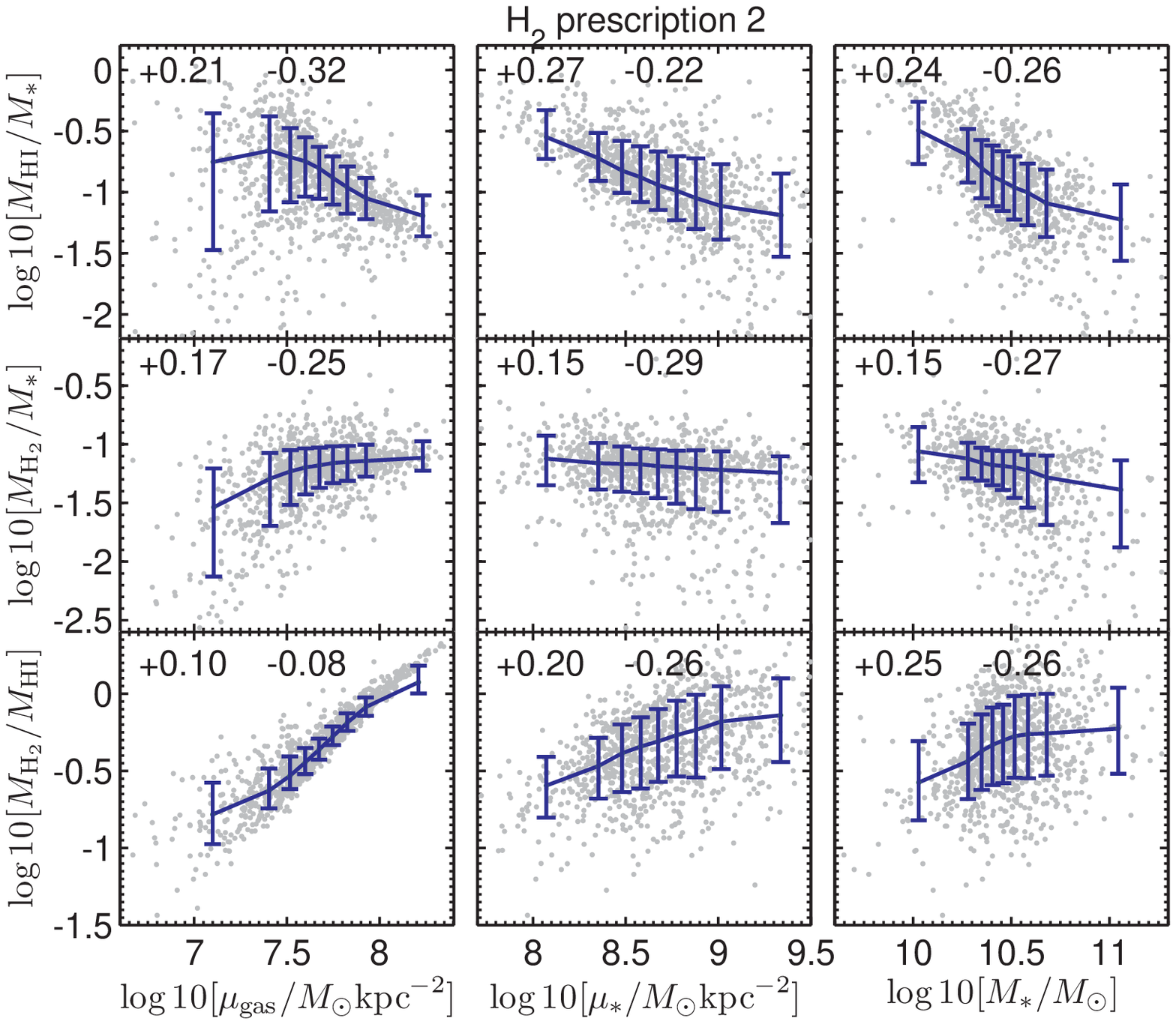}}
 \caption{ $\HIs$, $\Hs$ and $\HHI$ are plotted as a function of mass weighted mean gas surface density $\mugas$,
mean stellar surface density $\mu_*$, and stellar mass $M_*$ for model galaxies with $B/T \le 15$\% in dark matter
haloes with $v_{\rm{vir}}>$120km/s at z=0. Results are shown for both $\h2$ fraction prescriptions. In each plot, the
gray dots are the model galaxies and the blue curves show the mean values of $\HIs$, $\Hs$, $\HHI$ as a function of the
scale parameters on the x-axis. The error bars indicate the $\pm1\sigma$ scatter about the mean and the bins have been
chosen to contain an equal number of galaxies. We have also labelled each panel with the values of the mean scatter
about the relation in logarithmic units (the scatter towards higher values is denoted ``+'', and the scatter towards
lower values is denoted ``-'').} \label{model scale relation}
\end{figure*}

In Fig. \ref{model scale relation}, we plot $\HHI$, $\Hs$, $\HIs$ as a function of stellar mass $M_*$, mean stellar surface density
$\mu_*$ and mean gas surface density $\mugas$ for our model galaxies. In this analysis, we select the model galaxies with bulge
fraction $B/T \le 15\%$ in haloes with $v_{\rm{vir}}>$120km/s at z=0. These scaling relations should thus be regarded as appropriate
for massive late-type galaxies. We do not consider early-type galaxies or the dwarf galaxy population in this paper, because we do not
think our disk formation model is likely to provide an accurate representation of how such galaxies assemble.

We define $\mu_*$ as
\begin{equation}\label{eq:mustar}
\mu _*=\frac{0.5M_*}{\pi r_{50}^2},
\end{equation}
where $r_{50}$ is the radius enclosing half the stellar mass of the galaxy. Note that we have not implemented a detailed model for how
the stars in the bulge are distributed as a function of radius. In order to calculate $r_{50}$ in the models, we simply assume that
the stellar mass in the bulge is included within $r_{50}$. $\mugas$ is defined as the mass-weighted mean surface density of the gas in
the disk of the galaxy:
\begin{equation}\label{eq:mugas}
\mu _{{\rm{gas}}} = {m_{{\rm{gas}}}}/\pi \bar r_{{\rm{gas}}}^2,
\end{equation}
where the mass weighted mean radius of the gas disk $\bar r_{\rm{gas}}$ is defined as
\begin{equation}\label{eq:rgas}
\bar r_{\rm{gas}}=\frac{\int {rd{m_{\rm{gas}}}} }{\int {d{m_{\rm{gas}}}} }=\frac{1}{m_{\rm{gas}}}\sum\limits_i
{{r_i}{{\left(m_{\rm{gas}} \right)}_i}}.
\end{equation}
The sum in Equation (\ref{eq:rgas}) extends over the 30 rings used to represent the disk in our models.

The left and right panels in Fig. \ref{model scale relation} present results for the two $\h2$ fraction prescriptions. The gray dots
represent results for individual galaxies, while the blue curves with error bars indicate the mean value of the gas fraction (or
molecular-to-atomic ratio) as a function of the scale parameter on the x-axis. Error bars indicate the $\pm1\sigma$ scatter about the
mean. The atomic gas fraction $\HIs$ is inversely correlated with stellar mass, stellar surface mass density and gas mass surface
density. The molecular gas fraction $\Hs$ is correlated with the gas mass surface density, but shows little dependence on stellar mass
or stellar mass surface density. The molecular-to-atomic gas ratio $\HHI$ is tightly correlated with the surface mass density of the
gas, and more weakly correlated with stellar mass and stellar mass surface density.

A comparison of the two panels in Fig. \ref{model scale relation} shows that the two $\h2$ fraction prescriptions produce only
slightly different results. The most obvious difference is that the correlation between $\HHI$ and $\mugas$ is considerably tighter
for the pressure-based prescription than for the prescription based on the Krumholz et al. (2009) models. The very tight relation
between $\HHI$ and $\mugas$ in the right hand panel arises because $\Sigma_{\rm gas}$ dominates the contribution to the pressure for
the vast majority of the late-type galaxies in our simulation. However, galaxies of given gas surface density do have a spread in
metallicity, and this causes the increased scatter in the $\HHI$ versus $\mugas$ relation in the left panel.

\subsection{Physical origin of the scaling relations}

We will now attempt to elucidate the physical origin of the scaling relations discussed in the previous section. Because the
differences between the two $\h2$ fraction prescriptions are small, we will only focus on the results from prescription 1 in the
following analysis.

\begin{figure*}
\centering
  \subfigure[]{
  \includegraphics[angle=0,scale=0.52]{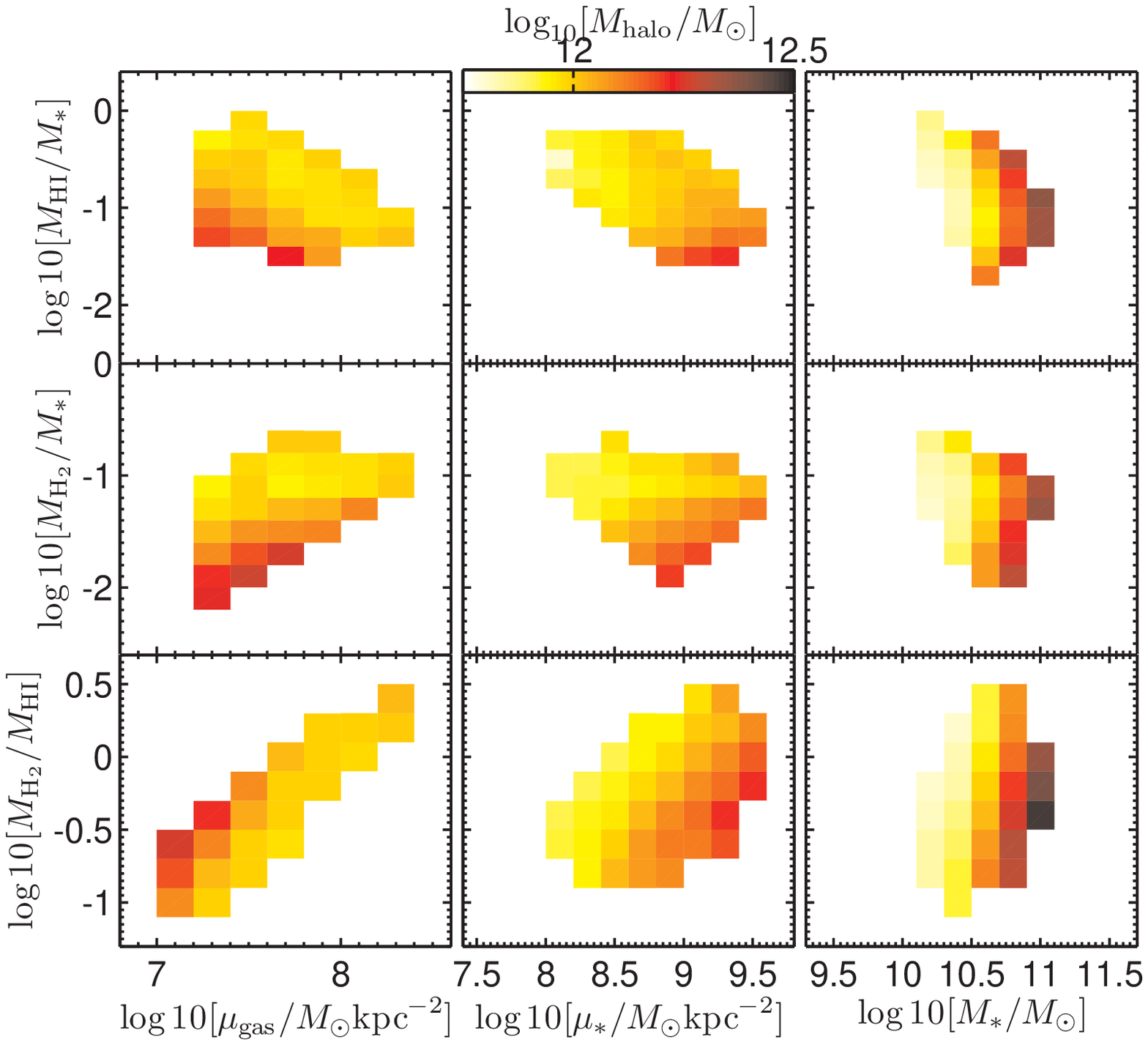}}
  \subfigure[]{
  \includegraphics[angle=0,scale=0.52]{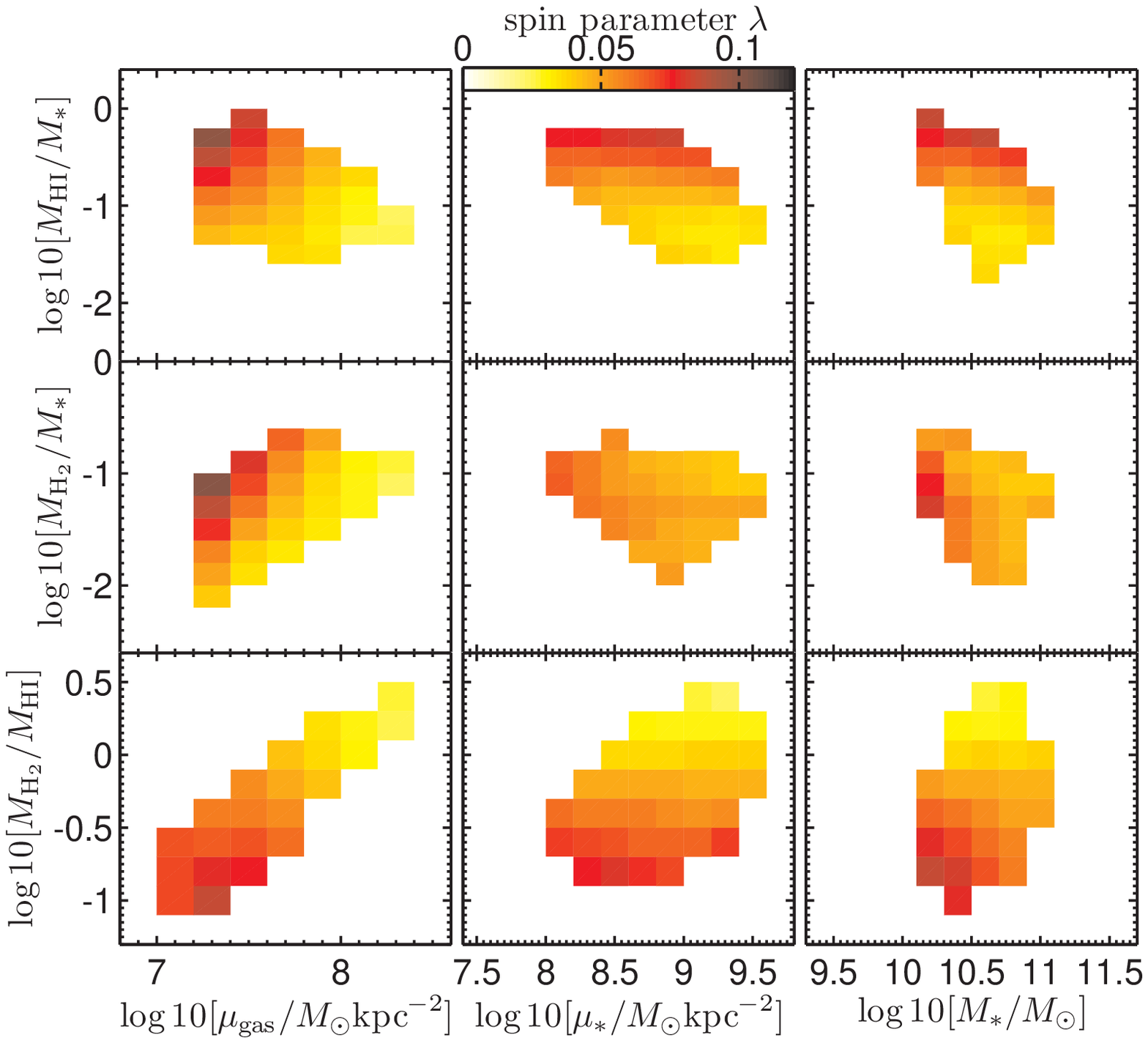}}\\
  \subfigure[]{
  \includegraphics[angle=0,scale=0.52]{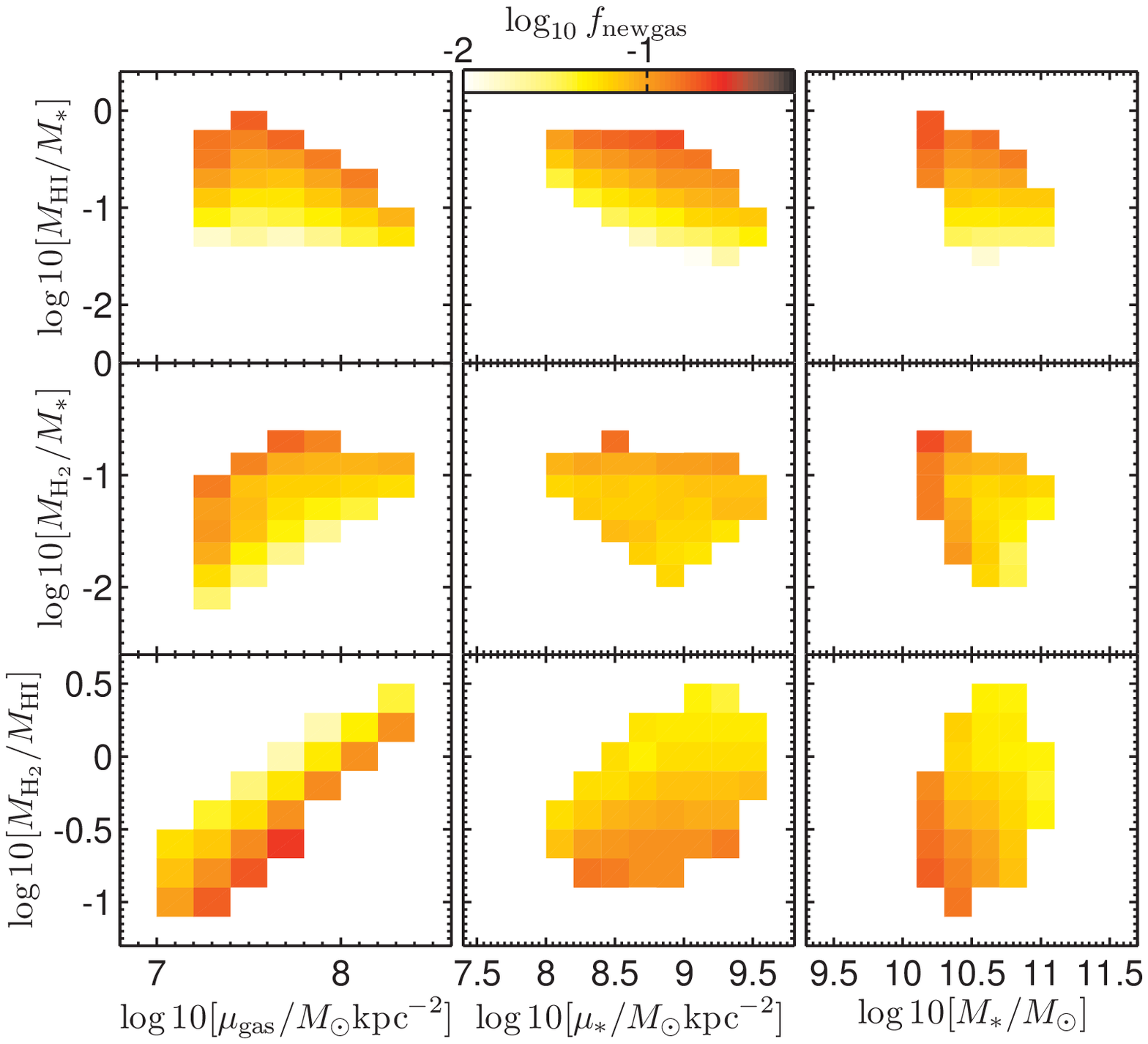}}
  \caption{The relations between $\HIs$, $\Hs$, $\HHI$ vs. mass weighted mean gas surface density $\mugas$, mean
  stellar surface density $\mu_*$, and stellar mass $M_*$ from model results for disk galaxies at redshift=0, which reside in
  haloes with virial velocities greater than 120 km/s. In each grid, the colours represent the mean values of halo virial mass
  $M_{\rm{vir}}$ (subfigure a), spin parameter $\lambda$ (subfigure b) and new gas fraction $f_{\rm{new}}$ (subfigure c).
  Each grid cell contains at least 100 galaxies.
 }\label{srgrid}
\end{figure*}

In our galaxy formation models, the mass and structure of the disk are set by parameters that cannot, in general, be directly
observed. These are:
\begin {enumerate}
\item The virial mass of the dark matter halo $M_{\rm{vir}}$.
\item The halo spin parameter $\lambda$ (Equation \ref{eq:lambda}).
\item The fraction of gas that has been accreted at recent epochs, $f_{\rm{new}}$.
\end {enumerate}
In this paper, we define $f_{\rm{new}}$ as the ratio of the mass of the gas that was accreted in the last Gigayear to the current
total disk (stellar+gas) mass of the galaxy disk:
\begin{equation}\label{eq:fnew}
f_{\rm{new}}=\frac{m_{\rm{cool}}(<1\rm{Gyr})}{m_{\rm{disk}}(z=0)}
\end{equation}

In Fig. \ref{srgrid}, we plot the same scaling relations as in the left panel of Fig. \ref{model scale relation}, but this time we
divide each plane into a set of cells, and we bin galaxies according to which cell they occupy. We then plot the average value of
$M_{\rm{vir}}$, $\lambda$ and $f_{\rm{new}}$ in each grid cell that is occupied by more than 100 galaxies. The results can be
summarized as follows:

(i) $M_{\rm{vir}}$: Panel (a) of Fig. \ref{srgrid} shows that more massive haloes host more massive galaxies with higher stellar
surface mass densities and lower gas fractions. The reason is that more baryons are available to form stars in more massive haloes. In
addition, feedback processes are less efficient at expelling baryons from the galaxy in massive haloes, so the {\em fraction} of the
baryons that cool and turn into stars is also higher.

(ii) $\lambda$: Panel (b) of \ref{srgrid} shows that galaxies in haloes with larger spin parameters have higher values of $\HIs$ and
$\Hs$, but lower values of $\HHI$. The spin parameter sets the contraction factor of the infalling gas. If it is large, a large
fraction of the recently accreted gas will be located in the extended, low density regions of the outer disk, where a smaller fraction
of the cold gas will be in molecular form. The resulting star formation rate surface densities will thus be low. This is why galaxies
in haloes with large spin parameter have low stellar surface densities, high gas fractions, but lower-than-average ratios of
molecular-to-atomic gas.

(iii) $f_{\rm{new}}$: Panel (c) of Fig. \ref{srgrid} shows that galaxies that have recently accreted a significant amount of new gas
have higher values of $\HIs$ and $\Hs$, but lower values of $\HHI$. Because galaxies form from the inside-out in our model, gas that
has been recently accreted will be located in the outer regions of the disks and have lower surface densities and lower molecular
fractions. To first order, therefore, the gas properties of galaxy that has experienced recent accretion will be similar to a galaxy
located in a halo with somewhat higher spin parameter.

The degeneracy between $\lambda$ and $f_{\rm{new}}$ can be analyzed in more detail by comparing panels (b) and (c). The basic trends
along the y-axis are quite similar for both panels: galaxies with higher-than-average gas fractions and lower-than average
molecular-to-atomic gas ratios could either have experienced a recent accretion event, or could simply have higher spin parameters.
Some notable differences do emerge, however, in the plots of gas fractions and molecular-to-atomic gas ratio as a function of
$\mugas$. Galaxies with $\mugas>10^8\ms\rm{kpc}^{-2}$ with higher-than-average gas fractions are very likely to have experienced a
recent gas accretion event. If one wishes to maximize the likelihood of identifying a galaxy that has experienced recent gas
accretion, Fig. \ref{srgrid} suggests that one should select: a) galaxies with high gas-to-star ratios and high gas surface densities,
b) galaxies with high gas-to-star ratios and low molecular-to-atomic gas ratios. For case (b), one would need to seek additional
evidence that the gas in the outer disk was accreted {\em recently}. We will discuss some ideas about how this might be done in the
final discussion.

The results in Fig. \ref{srgrid} can also help us understand the origin of the relations in Fig. \ref{model scale relation}. As seen
in Fig. \ref{srgrid}(a), more massive galaxies form in more massive haloes. Because feedback effects are less effective at preventing
the baryons in high mass haloes from cooling, a larger fraction of the available baryons is converted into stars, leading to galaxies
with higher stellar surface densities and lower gas-to-star ratios. This explains why $\HIs$ and $\Hs$ are anti-correlated both with
stellar mass and with stellar mass surface density. In Fig. \ref{srgrid}(c), we see that recent gas accretion rates are higher for low
mass galaxies. This is again because supernova feedback prevents baryons from cooling effectively in low mass haloes. Higher recent
gas infall rates lead to disks with lower molecular gas fractions. This explains why $\HHI$ is positively correlated with stellar mass
and stellar mass surface density. Because the trends in total gas-to-star ratio and molecular-to-atomic ratio work in opposite
directions, there are tighter correlations between the atomic gas fraction $\HIs$ and stellar mass and surface density, than between
the molecular gas fraction $\Hs$ and stellar mass and surface density.

Fig. \ref{model scale relation} shows that our models predict a very tight correlation between the atomic-to-molecular ratio and the
mean gas surface density of the disk. This is simply a reflection of our $\h2$ fraction prescriptions -- the local surface density of
the gas is the primary parameter that determines the fraction of gas that is converted to molecular form. Because galaxies with higher
values of $\mugas$ have higher total gas fractions {\em{and}} higher molecular-to-atomic gas ratios, there is a clear positive
correlation between $\Hs$ and $\mugas$. However, because more of the gas is transformed into molecular form at high total gas surface
densities, the relation between $\HIs$ and $\mugas$ is rather flat.

\subsection{Comparison with observations}

\begin{figure*}
\subfigure[]{
 \includegraphics[angle=-90,scale=0.52]{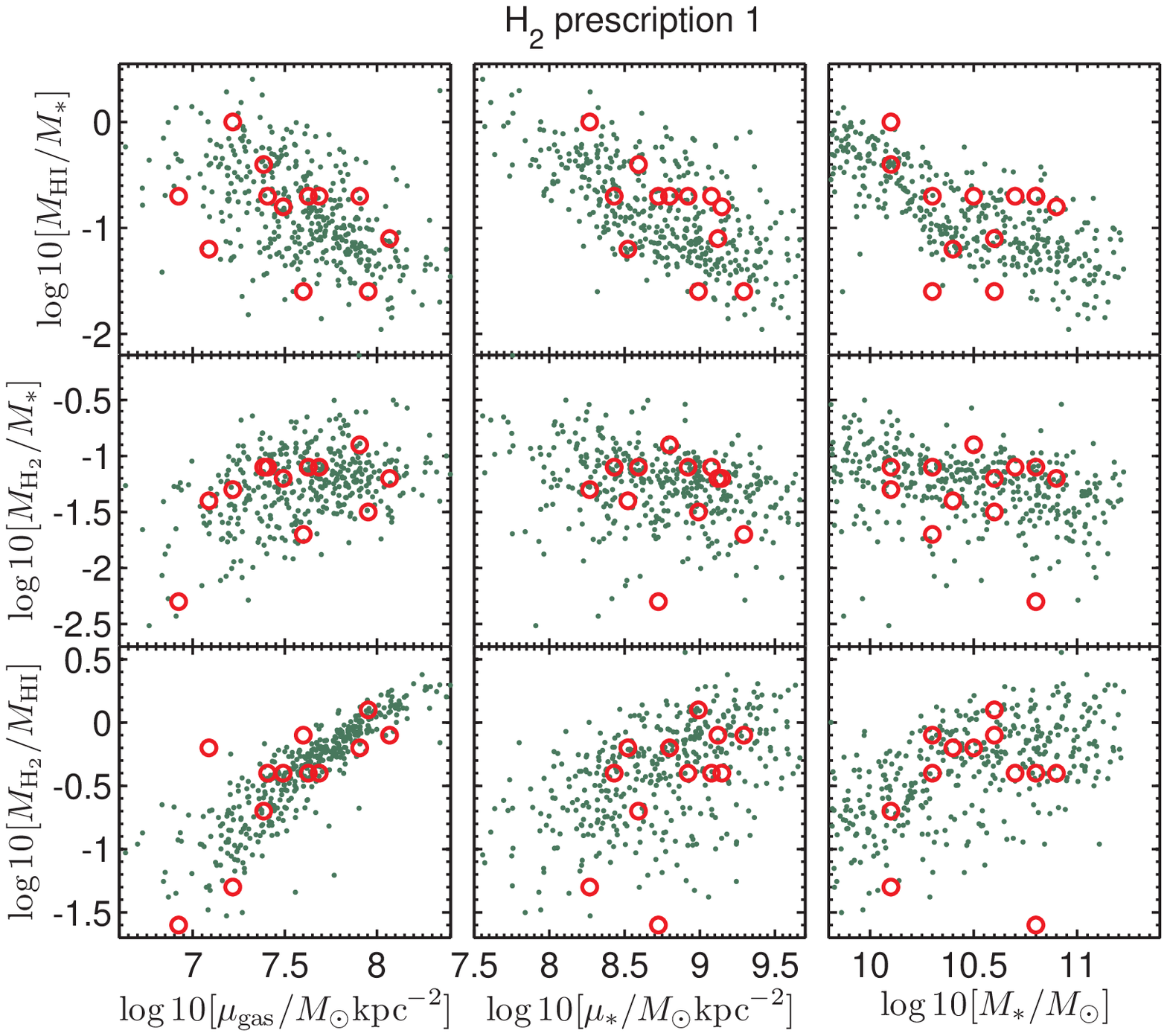}}
\subfigure[]{
 \includegraphics[angle=-90,scale=0.52]{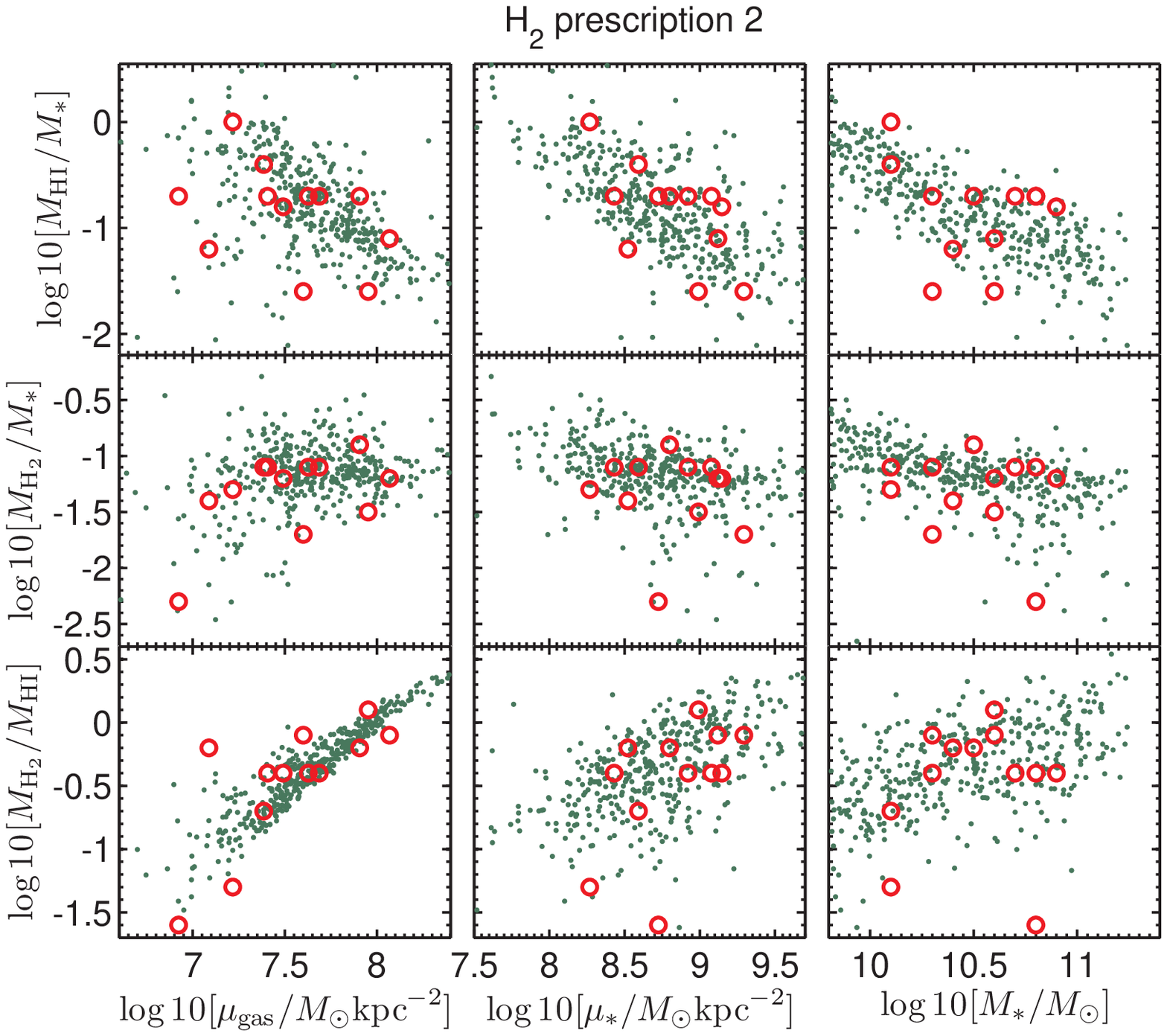}}
 \caption{Model relations between $\HHI$, $\Hs$, $\HIs$ and stellar mass $M_*$, mean stellar surface density $\mu_*$
 and mass weighted mean gas surface density $\mugas$ are compared with available observational data. The green dots
 are model results for disk galaxies at z=0 in dark matter haloes with $v_{\rm vir} > 120$ km/s. The red circles have
 been derived from the observational results tabulated in Leroy et al. (2008).
 }\label{scale relation}
\end{figure*}

In this section, we will compare the disk galaxy scaling relations predicted by our models with observations. One of the major
problems that we face is the lack of suitable data sets. Older data, for example as presented in the recent compilation by Obreschkow
\& Rawlings (2009), are beset by a variety of uncertainties and biases. The observed galaxies were selected using a variety of
different criteria, and single dish observations did not often cover the entire disk of the galaxy. These data sets are therefore
heterogeneous and the $\h2$ masses are likely to be inaccurate; it is therefore unclear whether the scaling relations derived from
them are useful for our purposes. For example, Figure 5 of Obreschkow \& Rawlings (2009) appears to indicate that local galaxies have
values of $\log(\HHI)$ that are distributed relatively uniformly from -1.5 to 1, i.e. if one believes these numbers, one would need to
understand why galaxies exhibit a range in molecular-to-atomic gas ratio of more than a factor of 100! In contrast, our models predict
a range in $\HHI$ closer to a factor of 10 (Fig. \ref{model scale relation}).

We turn once again to the small, but internally consistent THINGS/HERACLES data set. From the tables in Leroy et al. (2008), we derive
$\mu_*$ and $\mugas$ from the stellar and gas surface density profiles according to the definitions in Equation (\ref{eq:mustar}) \&
(\ref{eq:mugas}). The comparison between data and models presented in Fig. \ref{scale relation} is very encouraging. The gas fractions
and atomic-to-molecular gas ratios of the THINGS/HERACLES galaxies span the same range of values as our model galaxies. In addition,
one sees that: (1) The relation between $\HIs$ and $\mu_*$ and $m_*$ is steeper than that between $\Hs$ and $\mu_*$ and $m_*$, (2) The
relation between $\Hs$ vs $\mugas$ is tighter than that between $\HIs$ and $\mugas$, (3) $\HHI$ and $\mugas$ exhibit the strongest and
tightest correlations. These results are all consistent with our model predictions.

\begin{figure}
\centering
  \includegraphics[angle=-90,scale=0.52]{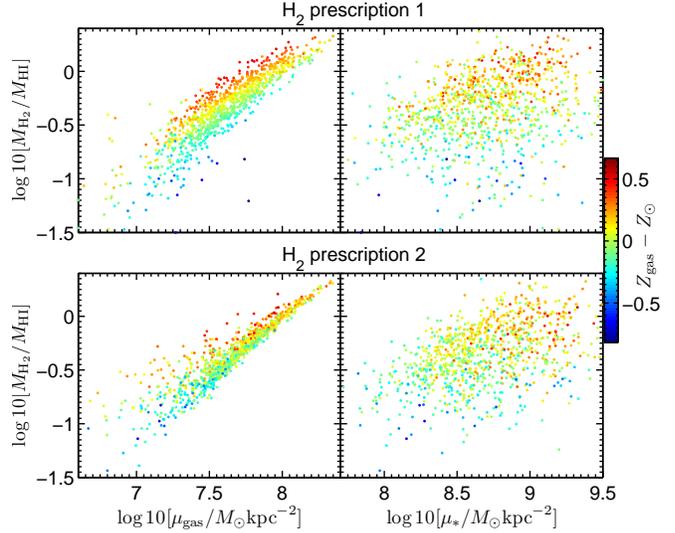}
  \caption{The molecular-to-atomic fraction $\HHI$ is plotted as a function of $\mugas$ and $\mu_*$
  for both $\h2$ fraction prescriptions, and the points are colour-coded according to gas-phase metallicity.
  The definition of the metallicity is $Z\equiv\log_{10}[M_{\rm{metal}}/M]$.
  }\label{scalingz}
\end{figure}

The current observations are not able to distinguish between the two $\h2$ fraction prescriptions. Adding more data points to Fig.
\ref{scale relation} may not help very much, because the main effect on the scaling relations is a change in the scatter in some of
the plots. One would have to understand the observational errors in considerable detail to know which part of the scatter was real.
Instead, we propose to find systematic effects in the scaling relations that arise as a consequence of the $\h2$ prescription. The
Krumholz et al. model predicts that much of the scatter arises as a result of metallicity differences between different galaxies. This
is illustrated in detail in Fig. \ref{scalingz}, where we plot the molecular-to-atomic fraction $\HHI$ as a function of $\mugas$ and
$\mu_*$, and we colour-code each galaxy according to its gas-phase metallicity. As can be seen, $\h2$ prescription 1 (the Krumholz et
al. model) predicts that there should be a clear stratification in metallicity at fixed $\mugas$ (and to a latter extent at fixed
$\mu_*$), with the most metal rich galaxies having higher values of $\HHI$. This is not seen for the pressure-based $\h2$
prescription. Since the gas-phase metallicity can be estimated using strong emission lines in optical spectra (e.g. Tremonti et al.
2004), this should be easily testable with future data sets, such as that provided by the CO Legacy database for GASS survey (see
http://www.mpa-garching.mpg.de/COLD$\_$GASS/).

\section{Summary and Discussion}

In this paper, we extend existing semi-analytic models to follow atomic and molecular gas in galaxies. We study how the condensed
baryons in present-day disk galaxies are partitioned between stars, HI and H$_2$ as a function of radius within the disk. Our new
model is implemented in the L-Galaxies semi-analytic code and is a modification of the models of Croton et al. (2006) and De Lucia \&
Blaizot (2007), in which dark matter halo merger trees derived from the Millennium Simulation form the ``skeleton'', on which we graft
simplified, but physically motivated, treatment of baryonic processes such as cooling, star formation, supernova feedback, and
chemical enrichment of the stars and gas. We fit these models to the radial surface density profiles of stars, HI and $\h2$ derived
from recent surveys of gas in nearby galaxies, making use of data from SINGS, THINGS, HERACLES and the BIMA SONG surveys.

We have used our models to explore how the relative mass fractions of atomic gas, molecular gas and stars are expected to vary as
a function of global galaxy scale parameters, including stellar mass, mean stellar surface density, and mean gas surface density. We
have attempted to elucidate how the trends can be understood in terms of the three variables that determine the partition of baryons
in disks: a) the mass of the dark matter halo, which determines the total mass of baryons that is able to cool and assemble in the
disk, b) the spin parameter of the halo, which sets the contraction factor of the gas, and thereby its surface density and molecular
fraction, c) the amount of gas that has been recently accreted from the external environment.

The main changes we have made to earlier models are the following:

(i) Each galactic disk is represented by a series of concentric rings. We assume that surface density profile of infalling gas in a
dark matter halo is exponential, with scale radius $r_{\rm d}$ that is proportional to the virial radius of the halo times the spin
parameter of the halo. As the Universe evolves, the dark matter halo grows in mass through mergers and accretion and the scale radius
of the infalling gas increases. Disk galaxies thus form from the inside out in our models. The ring representation allows us to track
the surface density {\em profiles} of the stars and gas as a function of cosmic time.

(ii) We include simple prescriptions for molecular gas formation processes in our models. We adopt two different ``recipes'': one
based on the analytic calculations by Krumholz et al. (2008), in which $\fh2$ is a function of the local surface density and
metallicity of the cold gas, and the other motivated by the work of Elmegreen (1989 \& 1993), Blitz \& Rosolowsky (2006), and
Obreschkow et al. (2009), in which the $\h2$ fraction is determined by the pressure of the ISM.

(iii) Motivated by the observational results of Leroy et al. (2008), we adopt a star formation law in which
$\Sigma_{\rm{SFR}}\propto\Sh2$ in the regime where the molecular gas dominates the total gas surface density, and
$\Sigma_{\rm{SFR}}\propto\sgas^2$ where atomic hydrogen dominates.

Our work leads to the following conclusions:

(i) A simple star formation law in which $\Sigma_{\rm{SFR}} \propto \Sigma_{\h2}$ leads to gas consumption time-scales
in the inner disk that are too short. In this paper, we simply patch over this problem by decreasing the efficiency of
supernova feedback in the inner disk.

(ii) The {\em mean} stellar, HI and $\h2$ surface density profiles of the disk galaxies in our model are only weakly
sensitive to the adopted $\h2$ fraction prescription. The reason for this is that for typical $L_*$ disk galaxies, the
local gas surface density is the main controlling parameter for both recipes. At low gas surface densities, the $\h2$
fraction depends sensitively on metallicity for the Krumholz et al. prescription, but considerably less sensitively on
stellar surface density $\mu_*$ for the pressure-based prescription. As a result, the correlation between
molecular-to-atomic fraction and $\mu_{\rm gas}$ for local disk galaxies exhibits more scatter if the Krumholz et al.
model is correct.

(iii) Our results indicate that galaxies that have recently accreted a significant amount of gas from the external environment are
characterized by higher-than-average {\em total} cold gas content. If the galaxy has high gas surface density, then this excess gas is
an unambiguous signature of a recent accretion event, because the time-scale over which gas is consumed into stars is short in such
systems. On the other hand, if the galaxy has low surface density, a higher-than-average total cold gas content could indicate a
recent accretion event, but it may also mean that the galaxy has a higher-than-average spin parameter. Higher spin parameters result
in disk galaxies with more extended distributions of cold gas, lower-than-average molecular-to-atomic ratios, and low star formation
efficiencies. For these ambiguous systems, one must seek additional evidence that the outer disks were assembled {\em recently}.

Although these conclusions are somewhat open-ended, they do suggest
avenues for further research.
We believe that a more realistic way forward
to solving the gas consumption timescale problem
would be to model radial inflow of the gas. Attempts have been
made to construct phenomenological models that do include radial mixing of the stars and gas in disks as well as the effect of this
mixing on the chemical evolution of the stars formed in the solar neighbourhood (e.g Sch{\"o}nrich \& Binney 2009). Results from
hydrodynamical simulations also indicate that the gas tends to flow inwards, while the stars migrate outwards (e.g Ro\v{s}kar et al.
2008). The main way to distinguish between different scenarios may be the predicted metallicity gradients. We intend to explore these
issues in more detail in future work.

In our model results, although the surface density profiles from the two $\h2$ fraction prescriptions are very similar, the models
indicate that one should, in principle, be able to confirm the metallicity-dependence of the molecular gas fraction predicted by the
Krumholz prescription, if one measures the average gas-phase metallicities of nearby disk galaxies using emission lines. Alternately,
one can break the degeneracy by observing systems where the metallicity is low but the pressure is high (Fumagalli, Krumholz \& Hunt
2010).

Another interesting issue is whether a galaxy's location in the gas scaling relation diagrams can serve as a diagnostic as to whether
it has accreted gas from the external environment. Although the theory of gas accretion in galaxies has received considerable
attention of late (e.g. Kere{\v s} et al. 2005; Dekel \& Birnboim 2006; Dekel et al. 2009), there is little {\em direct} observational
evidence that this occurs in practice. This is true both for galaxies in the local Universe and at high redshifts, where gas accretion
rates are expected to be much higher. Although average gas accretion rates are expected to be low at the present day, precise
quantification of the expected scaling relations for equilibrium disk galaxies may allow us to identify a subset of systems which
deviate significantly from the mean in terms of their gas content. Following the conclusion (iii), one may try to gain a better
understanding of the observationally detectable signatures of a recent gas accretion episode. Possible ways forward would be to look
for signatures of recent accretion in the observed age gradients of the stars or in the metallicity gradients of the gas in the disk.
One could also look for accretion signatures in the kinematics of the stars and the gas in the outer disks. Alternatively, one could
search for evidence of complex structure (e.g. tidal streams or shells) in the stellar haloes of gas-rich galaxies (Cooper at al
2010). We intend to explore these possibilities in more detail in future work.

Ongoing and future surveys, such as the Galex Arecibo Sloan Survey (GASS) (Catinella et al. 2010) and the COLD GASS survey carried out
at the IRAM 30m telescope (Saintonge et al. in preparation) will enable us to quantify the scaling relations discussed in this paper
in considerable detail. These surveys will provide interesting targets for follow-up programs, which may help us understand that
extent to which galaxies still accrete gas at the present day. In the next few years, it will become possible to observe gas in
galaxies at higher redshifts using facilities such as ALMA and Square Kilometer Array pathfinder experiments such as ASKAP or MEERKAT.
We are certain that our simplified treatment of disk formation in concentric rings that undergo no radial mixing will not be a good
way to describe the assembly of the clumpy, highly turbulent disks that are now known to exist at $z \sim 2$ (e.g. Genzel et al.
2008). Nevertheless, we believe that our models may still be useful in elucidating the gaseous and chemical evolution of disks over a
somewhat smaller range in lookback time.

\section*{Acknowledgments}
Jian Fu thanks the Joint Program between Chinese Academic of Sciences and Max-Planck-Gesellschaft for the chance of visiting the
Max-Planck-Institut f\"ur Astrophysik, and Deutsche Forschungsgemeinschaft for the extra support. He also thanks the support from the
National Science Foundation of the Key Project No. 10833005, the Group Innovation Project No. 10821302, and by 973 program No.
2007CB815402. Guinevere Kauffmann thanks Reinhard Genzel for conversations that inspired this work. Mark Krumholz acknowledges support
from the Alfred P. Sloan Foundation, the National Science Foundation through grant AST-0807739, NASA through ATFP grant NNX09AK31G,
and NASA through the Spitzer Space Telescope Theoretical Research Program, provided by a contract issued by the Jet Propulsion
Laboratory.

\def\apj{ApJ}
\def\apjl{ApJL}
\def\apjs{ApJS}
\def\aj{AJ}
\def\aap{A\&A}
\def\araa{ARA\&A}
\def\aapss{A\&AS}
\def\mnras{MNRAS}
\def\nature{Nature}
\def\apss{Ap\&SS}
\def\pasp{PASP}

{}


\begin{thebibliography}{}

\bibitem[\protect\citeauthoryear{Arnett, David}{1996}]{} Arnett, David (1996). Supernovae and Nucleosynthesis (First edition ed.). Princeton, New Jersey: Princeton University Press.
\bibitem[\protect\citeauthoryear{Avila-Reese, Firmani, \& Hern{\'a}ndez}{1998}]{} Avila-Reese V., Firmani C., Hern{\'a}ndez X., 1998, ApJ, 505, 37
\bibitem[\protect\citeauthoryear{Bigiel et al.}{2008}]{} Bigiel, F., Leroy, A., Walter, F., Brinks, E., de Blok, W.~J.~G., Madore, B., \& Thornley, M.~D.\ 2008, \aj, 136, 2846
\bibitem[\protect\citeauthoryear{Blitz \& Rosolowsky}{2004}]{} Blitz, L., \& Rosolowsky, E.\ 2004, \apjl, 612, L29
\bibitem[\protect\citeauthoryear{Blitz \& Rosolowsky}{2006}]{} Blitz, L., \& Rosolowsky, E.\ 2006, \apj, 650, 933
\bibitem[\protect\citeauthoryear{Boissier \& Prantzos}{1999}]{} Boissier, S., \& Prantzos, N.\ 1999, \mnras, 307, 857
\bibitem[\protect\citeauthoryear{Boselli, Lequeux, \& Gavazzi}{2002}]{} Boselli A., Lequeux J., Gavazzi G., 2002, Ap\&SS, 281, 127
\bibitem[\protect\citeauthoryear{Bottema}{1993}]{} Bottema, R.\ 1993, \aap, 275, 16
\bibitem[\protect\citeauthoryear{Boulanger \& Viallefond}{1992}]{} Boulanger, F., \& Viallefond, F.\ 1992, \aap, 266, 37
\bibitem[\protect\citeauthoryear{Bower et al.}{2006}]{} Bower, R.~G., Benson, A.~J., Malbon, R., Helly, J.~C., Frenk, C.~S., Baugh, C.~M., Cole, S., \& Lacey, C.~G.\ 2006, \mnras, 370, 645
\bibitem[\protect\citeauthoryear{Catinella et al.}{2010}]{} Catinella, B., Schiminovich, D., Kauffmann, G., et al.\ 2010, \mnras, 403, 683
\bibitem[\protect\citeauthoryear{Chabrier}{2003}]{} Chabrier, G.\ 2003, \pasp, 115, 763
\bibitem[\protect\citeauthoryear{Cole et al.}{1994}]{} Cole, S., Aragon-Salamanca, A., Frenk, C.~S., Navarro, J.~F., \& Zepf, S.~E.\ 1994, \mnras, 271, 781
\bibitem[\protect\citeauthoryear{Cooper et al.}{2010}]{} Cooper, A.~P., Cole, S., Frenk, C. S., et al.\ 2010, \mnras, 756
\bibitem[\protect\citeauthoryear{Croton et al.}{2006}]{} Croton, D.~J., Springel, V., White, S. D. M., et al. \ 2006, \mnras, 365, 11
\bibitem[\protect\citeauthoryear{Dalcanton, Spergel, \& Summers }{1997}]{} Dalcanton J.~J., Spergel D.~N., Summers F.~J., 1997, ApJ, 482, 659
\bibitem[\protect\citeauthoryear{Dekel \& Birnboim}{2006}]{} Dekel, A., \& Birnboim, Y.\ 2006, \mnras, 368, 2
\bibitem[\protect\citeauthoryear{Dekel et al.}{2009}]{} Dekel, A., Birnboim, Y., Engel, G., et al.\ 2009, \nature, 457, 451
\bibitem[\protect\citeauthoryear{De Lucia \& Blaizot}{2007}]{} De Lucia, G., \& Blaizot, J.\ 2007, \mnras, 375, 2
\bibitem[\protect\citeauthoryear{Dutton}{2009}]{} Dutton, A.~A.\ 2009, \mnras, 396, 121
\bibitem[\protect\citeauthoryear{Dutton \& van den Bosch}{2009}]{} Dutton, A.~A., \& van den Bosch, F.~C.\ 2009, \mnras, 396, 141
\bibitem[\protect\citeauthoryear{Elmegreen}{1989}]{} Elmegreen, B.~G.\ 1989, \apj, 338, 178
\bibitem[\protect\citeauthoryear{Elmegreen}{1993}]{} Elmegreen, B.~G.\ 1993, \apj, 411, 170
\bibitem[\protect\citeauthoryear{Fu et al.}{2009}]{} Fu, J., Hou, J.~L., Yin, J., \& Chang, R.~X.\ 2009, \apj, 696, 668
\bibitem[\protect\citeauthoryear{Fumagalli et al. }{2010}]{} Fumagalli, M., Krumholz, M. R., \& Hunt, L. K. 2010, \apj, submitted
\bibitem[\protect\citeauthoryear{Genzel et al.}{2008}]{} Genzel, R., Burkert, A., Bouch{\'e}, N., et al.\ 2008, \apj, 687, 59
\bibitem[\protect\citeauthoryear{Gnedin et al.}{2009}]{} Gnedin, N.~Y., Tassis,K., \& Kravtsov, A.~V.\ 2009, \apj, 697, 55
\bibitem[\protect\citeauthoryear{Guo et al.}{2010}]{} Guo, Q., White, S., Li, C., \& Boylan-Kolchin, M.\ 2010, \mnras, 367
\bibitem[\protect\citeauthoryear{Helfer et al.}{2003}]{} Helfer, T.~T., Thornley, M.~D., Regan, M.~W., Wong, T., Sheth, K., Vogel, S.~N., Blitz, L., \& Bock, D.~C.-J.\ 2003, \apjs, 145, 259
\bibitem[\protect\citeauthoryear{Laurikainen et al.}{2007}]{} Laurikainen, E., Salo, H., Buta, R., \& Knapen, J.~H.\ 2007, \mnras, 381, 401
\bibitem[\protect\citeauthoryear{Kauffmann et al.}{1993}]{} Kauffmann, G., White, S.~D.~M., \& Guiderdoni, B.\ 1993, \mnras, 264, 201
\bibitem[\protect\citeauthoryear{Kauffmann}{1996}]{} Kauffmann, G.\ 1996, \mnras, 281, 475
\bibitem[\protect\citeauthoryear{Kauffmann et al.}{1999}]{} Kauffmann, G., Colberg, J.~M., Diaferio, A., \& White, S.~D.~M.\ 1999, \mnras, 307, 529
\bibitem[\protect\citeauthoryear{Kennicutt}{1989}]{} Kennicutt, R.~C., Jr.\ 1989, \apj, 344, 685
\bibitem[\protect\citeauthoryear{Kennicutt}{1998}]{} Kennicutt, R.~C., Jr.\ 1998, \apj, 498, 541
\bibitem[\protect\citeauthoryear{Kennicutt et al.}{2003}]{} Kennicutt, R.~C., Jr., Armus, L., Bendo, G., et al.\ 2003, \pasp, 115, 928
\bibitem[\protect\citeauthoryear{Keres et al.}{2003}]{} Keres, D., Yun, M.~S., \& Young, J.~S.\ 2003, \apj, 582, 659
\bibitem[\protect\citeauthoryear{Kere{\v s} et al.}{2005}]{} Kere{\v s}, D., Katz, N., Weinberg, D.~H., \& Dav{\'e}, R.\ 2005, \mnras, 363, 2
\bibitem[\protect\citeauthoryear{Krumholz et al.}{2008}]{} Krumholz, M.~R., McKee, C.~F., \& Tumlinson, J.\ 2008, \apj, 689, 865
\bibitem[\protect\citeauthoryear{Krumholz et al.}{2009}]{} Krumholz, M.~R., McKee, C.~F., \& Tumlinson, J.\ 2009, \apj, 693, 216
\bibitem[\protect\citeauthoryear{Leroy et al.}{2008}]{} Leroy, A.~K., Walter, F., Brinks, E., Bigiel, F., de Blok, W.~J.~G., Madore, B., \& Thornley, M.~D.\ 2008, \aj, 136, 2782
\bibitem[\protect\citeauthoryear{Li \& White}{2009}]{} Li, C., \& White, S.~D.~M.\ 2009, \mnras, 398, 2177
\bibitem[\protect\citeauthoryear{Madau et al.}{1998}]{} Madau, P., Pozzetti, L., \& Dickinson, M.\ 1998, \apj, 498, 106
\bibitem[\protect\citeauthoryear{McKee \& Krumholz}{2010}]{} McKee, C.~F., \& Krumholz, M.~R.\ 2010, \apj, 709, 308
\bibitem[\protect\citeauthoryear{Mo et al.}{1998}]{} Mo, H.~J., Mao, S., \& White, S.~D.~M.\ 1998, \mnras, 295, 319
\bibitem[\protect\citeauthoryear{Obreschkow et al.}{2009}]{} Obreschkow, D., Croton, D., DeLucia, G., Khochfar, S., \& Rawlings, S.\ 2009, \apj, 698, 1467
\bibitem[\protect\citeauthoryear{Obreschkow \& Rawlings}{2009}]{} Obreschkow, D., \& Rawlings, S.\ 2009, \mnras, 394, 1857
\bibitem[\protect\citeauthoryear{Reynolds}{2004}]{} Reynolds, R.~J.\ 2004, Advances in Space Research, 34, 27
\bibitem[\protect\citeauthoryear{Ro{\v s}kar et al.}{2008}]{} Ro{\v s}kar, R., Debattista, V.~P., Quinn, T.~R., Stinson, G.~S., \& Wadsley, J.\ 2008, \apjl, 684, L79
\bibitem[\protect\citeauthoryear{Saintonge et al. in preparation }{2010}]{} Saintonge, A., et al. 2010 in preparation
\bibitem[\protect\citeauthoryear{Schmidt}{1959}]{} Schmidt, M.\ 1959, \apj, 129, 243
\bibitem[\protect\citeauthoryear{Sch{\"o}nrich \& Binney}{2009}]{} Sch{\"o}nrich, R., \& Binney, J.\ 2009, \mnras, 396, 203
\bibitem[\protect\citeauthoryear{Somerville \& Primack}{1999}]{} Somerville, R.~S., \& Primack, J.~R.\ 1999, \mnras, 310, 1087
\bibitem[\protect\citeauthoryear{Somerville et al.}{2001}]{} Somerville, R.~S., Primack, J.~R., \& Faber, S.~M.\ 2001, \mnras, 320, 504
\bibitem[\protect\citeauthoryear{Springel et al. }{2005}]{} Springel V., White, S. D. M., Jenkins, A., et al., 2005, Natur, 435, 629
\bibitem[\protect\citeauthoryear{Sutherland \& Dopita}{1993}]{} Sutherland, R.~S., \& Dopita, M.~A.\ 1993, \apjs, 88, 253
\bibitem[\protect\citeauthoryear{Toomre}{1964}]{} Toomre, A.\ 1964, \apj, 139, 1217
\bibitem[\protect\citeauthoryear{Tremonti et al. }{2004}]{} Tremonti C.~A., Heckman, T. M., Kauffmann, G., et al., 2004, ApJ, 613, 898
\bibitem[\protect\citeauthoryear{van den Bosch }{1998}]{} van den Bosch F.~C., 1998, ApJ, 507, 601
\bibitem[\protect\citeauthoryear{Walter et al.}{2008}]{} Walter, F., Brinks, E., de Blok, W.~J.~G., Bigiel, F., Kennicutt, R.~C., Thornley, M.~D., \& Leroy, A.\ 2008, \aj, 136, 2563
\bibitem[\protect\citeauthoryear{Weinzirl et al.}{2009}]{} Weinzirl, T., Jogee, S., Khochfar, S., Burkert, A., \& Kormendy, J.\ 2009, \apj, 696, 411
\bibitem[\protect\citeauthoryear{Wong \& Blitz}{2002}]{} Wong, T., \& Blitz, L.\ 2002, \apj, 569, 157
\bibitem[\protect\citeauthoryear{White \& Frenk}{1991}]{} White, S.~D.~M., \& Frenk, C.~S.\ 1991, \apj, 379, 52
\bibitem[\protect\citeauthoryear{Zwaan et al.}{2005}]{} Zwaan, M.~A., Meyer, M.~J., Staveley-Smith, L., \& Webster, R.~L.\ 2005, \mnras, 359, L30

\end{thebibliography}
\end{document}